\documentclass[aps,prd,twocolumn,unsortedaddress,showpacs,nofootinbib]{revtex4-1}

\usepackage{amsmath,amssymb,graphicx}

\begin{document}

\title{Probing a dark matter density spike at the Galactic Center}

\author{Thomas Lacroix}
\affiliation{UMR7095, Institut d'Astrophysique de Paris, 98 bis boulevard Arago, 75014 Paris, France}
\email{lacroix@iap.fr}
\author{C\'{e}line B\oe hm}
\affiliation{Institute for Particle Physics Phenomenology, Durham University, Durham, DH1 3LE, United Kingdom}
\affiliation{LAPTH, U. de Savoie, CNRS, BP 110, 74941 Annecy-Le-Vieux, France}
\email{c.m.boehm@durham.ac.uk}
\author{Joseph Silk}
\affiliation{UMR7095, Institut d'Astrophysique de Paris, 98 bis boulevard Arago, 75014 Paris, France}
\affiliation{The Johns Hopkins University, Department of Physics and Astronomy,
3400 N. Charles Street, Baltimore, Maryland 21218, USA}
\affiliation{Beecroft Institute of Particle Astrophysics and Cosmology, Department of Physics,
University of Oxford, Denys Wilkinson Building, 1 Keble Road, Oxford OX1 3RH, United Kingdom}
\email{silk@iap.fr}

\date{\today}

\begin{abstract}
The dark matter halo profile in the inner Galaxy is very uncertain. Yet its radial dependence toward the Galactic Center is of crucial importance for the determination of the gamma-ray and radio fluxes originating from dark matter annihilations. Here we use synchrotron emission to probe the dark matter energy distribution in the inner Galaxy. We first solve the problem of the cosmic ray diffusion on very small scales, typically smaller than $ 10^{-3}\ \rm pc $, by using a Green's function approach and use this technique to quantify the effect of a spiky profile ($\rho(r) \propto r^{-7/3}$) on the morphology and intensity of the synchrotron emission expected from dark matter. We illustrate our results using 10 and 800 GeV candidate weakly interacting dark matter particles annihilating directly into $e^+ e^-$. Our most critical assumptions are that the dark matter is heavier than a few GeV and directly produces a reasonable amount of electrons and positrons in the Galaxy. We conclude that dark matter indirect detection techniques (including the Planck experiment) could be used to shed light on the dark matter halo profile on scales that lie beyond the capability of any current numerical simulations.
\end{abstract}

\pacs{95.35.+d, 96.50.S-, 98.35.Jk}

\maketitle

\section{Introduction}

In the framework of cold dark matter ($ \mathrm{\Lambda CDM} $), dark matter comprises about $ 27 \% $ of the energy content of the universe. Consequently, unveiling the nature of dark matter (DM) is one of the greatest challenges of modern cosmology. The popular solution to accommodate several astrophysical and cosmological observations is to assume that DM is made of weakly interacting massive particles (WIMPs), as predicted in, e.g., supersymmetric extensions of the standard model of particle physics. However, the properties of such particles are unknown at present and need to be determined. In this context, indirect detection can provide constraints that are complementary to direct detection experiments, as well as accelerator and collider physics probes. 

Given that the annihilation rate scales with the square of the DM density, the Galactic Center (GC)---where the DM density is expected to be highest---is a promising region for such indirect searches \cite{JoeSalati,BergstromUlio,BEnssS,multi-wavelength,Bringmann}. Indeed the authors of Ref.~\cite{GordonMacias} find that the GC provides stronger constraints than dwarf galaxies on the DM annihilation cross section.

However, the DM halo profile toward the center, i.e., at small radii (sub-kpc) is unknown. In Ref~\cite{GondoloSilk} it was suggested that there could be a strong enhancement of the DM energy density (referred to as a``spike") around the supermassive black hole Sgr A* at the GC, but this remains to be established. The effects of annihilations and especially of dynamical relaxation by stellar interactions may soften this spike \cite{DM_dynamics_GC}, but the full range of dynamical effects has not been fully explored. For example, the competing effects of the black hole growth time scale, the adiabatic response of the dark matter, and the core relaxation time by stellar dynamical heating are of the same order of magnitude. In this paper we will therefore focus on a range of possible very dense inner spike profiles and their effects on the synchrotron emission originating from the DM. This will enable us to determine whether present experiments can constrain the DM distribution very near to the GC. 

Synchrotron emission critically relies on cosmic ray propagation, but cosmic ray diffusion at very small scales requires a specific technique that has not been presented before in the literature. This method relies on a careful treatment of the Green's functions by adapting the integration step to three different regimes defined in terms of the distance from the GC.

To illustrate our point, we will focus on $ 10\ \rm GeV $ DM particles but will also consider heavy (e.g.~$ 800\ \rm GeV $) DM candidates for the sake of completeness. At present, there are claims of possible evidence for light annihilating DM particles \cite{BEnssS} in direct detection experiments (notably DAMA/LIBRA \cite{DAMA/LIBRA}, CoGeNT \cite{cogent}, and CDMS \cite{CDMS}), but there are also contradictory signals \cite{XENON,LUX}. There are in addition constraints from radio signatures in Galaxy clusters and the center of the Milky Way \cite{BEnssS,BEnssS2,Synchrotron_wimps}, as well as in off-center regions of the Milky Way \cite{Bryan}.\footnote{There is also a constraint from the positron flux in the Galaxy \cite{Hooper_AMS} using positron data, but any case for actual detection of an annihilation signal assumes an excellent knowledge of the astrophysical backgrounds, which is questionable.} Such particles may nevertheless provide us with a possible explanation for the nonthermal radio filaments observed at the GC \cite{Hooper_radio_filaments} and are worth considering even if it is just for illustrative purposes.

In Sec.~\ref{CR propagation}, we recall the general framework of cosmic-ray propagation and describe the technique that we use to model the diffusion of electrons produced by DM on very small scales and their subsequent synchrotron emission. In Sec.~\ref{results}, we show the effect of a DM halo profile with a spike on the synchrotron flux and discuss the potential for observations.

\section{Propagation of cosmic rays and synchrotron emission}
\label{CR propagation}

In this section we revisit the propagation of electrons and positrons originating from DM in light of the technique that we use to solve the diffusion problem at very small scales, and compute the resulting synchrotron emission from the inner region of the Milky Way.

\subsection{Transport equation}

After their injection by DM, electrons and positrons propagate in the Galaxy following the diffusion-loss equation of cosmic rays. Assuming a steady state, this equation reads \cite{Longair, Itilda, Synchrotron_wimps}
\begin{equation}
\label{transport equation}
K \nabla^{2}\psi + \dfrac{\partial}{\partial E}(b\psi) + q = 0,
\end{equation}
where $ \psi \equiv \psi ( \vec{x},E) $ is the particle spectrum (number density per unit energy) at location $ \vec{x} $ and energy $ E $. $ \nabla^{2} $ is the Laplacian operator, $q \equiv q(\vec{x},E) $ is the source term, and $ b(\vec{x},E) $ describes the total energy loss of the particle. The diffusion coefficient $ K $ models the transport through the Galactic magnetic field. It is assumed to be independent of the position of the cosmic rays and is generally parametrized in the following way \cite{Itilda, Synchrotron_wimps, Synchrotron_DM_decay}: $ K(E) = K_{0} \left( E/E_{0} \right) ^{\delta} $, where $ E_{0} $ is an energy normalization taken to be $ 1 \ \rm GeV $.

Cosmic rays in the Milky Way Galaxy are confined by the Galactic magnetic field to a diffusion zone modelled by a cylinder of radius $ R_{\mathrm{gal}} = 20 \ \rm kpc $ and half-thickness $ L $ (defined with respect to the Galactic plane). Three parameters therefore govern the propagation of cosmic rays in this diffusion zone: the half-thickness $ L $, the normalization of the diffusion coefficient $ K_{0} $, and its energy dependence $ \delta $. The best fit to the cosmic-ray measurements of the boron-to-carbon (B/C) ratio at Earth's position \cite{Synchrotron_wimps} is referred to as the medium (MED) parameter set. In this work, we extrapolate the value of the propagation parameters obtained at Earth's position all the way down to the GC. The two other sets of propagation parameters, the so-called minimum (MIN) and maximum (MAX) sets, correspond to the minimal and maximal primary antiproton fluxes which are compatible with the B/C analysis \cite{Synchrotron_wimps}. The three sets of parameters are given by
\begin{align}
\label{propa_sets}
\mathrm{MIN}:\ L & = 1\ \mathrm{kpc},K_{0} = 0.0016\ \rm kpc^{2}\ Myr^{-1},\delta = 0.85, \nonumber \\ 
\mathrm{MED}:\ L & = 4 \ \mathrm{kpc},K_{0} = 0.0112 \ \rm kpc^{2} \ Myr^{-1},\delta = 0.7, \nonumber \\
\mathrm{MAX}:\ L & = 15 \ \mathrm{kpc},K_{0} = 0.0765 \ \rm kpc^{2} \ Myr^{-1},\delta = 0.46.
\end{align}
Consequently, the MIN and MAX sets allow one to quantify the uncertainties on the diffusion models compatible with observational data.

\subsection{Source term}
\label{Dark matter spike}

In this work we assume that DM annihilates directly into electrons and positrons and that no other source can produce electrons and positrons. As a result the source term reads
\begin{equation}
\label{source term}
q( \vec{x},E) = \dfrac{1}{2} \left\langle \sigma v \right\rangle \left( \dfrac{\rho ( \vec{x})}{m_{\mathrm{DM}}} \right) ^{2} \dfrac{\mathrm{d}n}{\mathrm{d}E}(E),
\end{equation}
where $ \left\langle \sigma v \right\rangle $ is the thermally averaged cross section times relative velocity of the DM particles, $ \rho(\vec{x})$ is the DM density at position $\vec{x}$, $ m_{\mathrm{DM}} $ is the mass of the DM particles, and the numerical factor $ 1/2 $ arises when assuming that the DM particles are self-conjugate (e.g., Majorana particles). We take this value in the following, but for non-self-conjugate DM (e.g., Dirac particles), this factor becomes $ 1/4 $. The term $ \mathrm{d}n/\mathrm{d}E $ is the energy spectrum of the electrons and positrons for a single annihilation. In our case the electron and positron energy distribution can be described by a Dirac function $ \mathrm{d}n/\mathrm{d}E = \delta (E - m_{\mathrm{DM}}) $, due to the kinematics of the DM pair annihilation process into $e^+ e^-$.

To go one step further, we need to specify the DM energy distribution $\rho(\vec{x})$ in the Galaxy. We will consider two types of DM halo profiles: a Navarro-Frenk-White (NFW) \cite{NFW} and a NFW+spike profile. As the DM energy density for such profiles is divergent toward the GC, we need to specify a prescription (cutoff scale) to avoid getting unphysical results. Although such a prescription is in principle required for a NFW profile, it was shown in Ref.~\cite{BoehmLavalle} that the resolution of the instrument actually regularized the divergence. Such a regularization cannot be used in the case of spiky profiles because the increase in the DM density toward the center is too steep. We thus introduce the notion of saturation density $ \rho_{\mathrm{sat}} $ that defines a plateau distribution (i.e., a core) at any scale $ r < r_{\rm{sat}} $, with $ r_{\rm{sat}} $ the saturation radius defined by the equality $\rho(r_{\rm{sat}}) = \rho_{\rm{sat}}$. A natural value for $ \rho_{\mathrm{sat}} $ is given by the saturation density set by annihilations $ \rho_{\mathrm{sat}}^{\rm{ann}} $, which corresponds to the equality between the annihilation characteristic time and the infall time $ t_{\mathrm{i}} $ of DM particles onto the central black hole:
\begin{equation}
\rho_{\mathrm{sat}}^{\rm{ann}} = \dfrac{m_{\mathrm{DM}}}{\left\langle \sigma v \right\rangle t_{\mathrm{i}}}.
\label{rhosat}
\end{equation}
We assume a conservative value of the infall time, $ t_{\mathrm{i}} = 10^{10}\ \rm yr $, by taking it to be equal to the age of the black hole, as in Ref.~\cite{Merritt}. For the NFW$+$spike profile, we will thus assume the following radial dependence:
\begin{equation}
\rho (r) =
\begin{cases}
\rho_{\odot} \dfrac{r_{\odot}}{r} \left( \dfrac{1 + r_{\odot}/r_{\mathrm{s}}}{1 + r/r_{\mathrm{s}}} \right) ^{2} & r > R_{\mathrm{spike}} \\
\rho_{\mathrm{sat}} \left( \dfrac{r}{r_{\mathrm{sat}}} \right) ^{-\gamma_{\mathrm{spike}}} & r_{\mathrm{sat}} < r \leqslant R_{\mathrm{spike}} \\
\rho_{\mathrm{sat}} & r \leqslant r_{\mathrm{sat}},
\end{cases}
\end{equation}
where $ \rho_{\odot} = 0.3\ \rm GeV\ cm^{-3} $ is the local DM density at the Sun's position, $ r_{\odot} = 8.5\ \rm kpc$; $ R_{\mathrm{spike}} $ is the radius of the spike; and $ r_{\mathrm{s}} = 20\ \rm kpc $ parametrizes the NFW profile. The value of the index $ \gamma_{\mathrm{spike}} $ is expected to lie between 2.25 and 2.5 as suggested in Ref.~\cite{Gondolo}. When the values of the DM mass or annihilation cross section are changed, the very inner part of the density profile is changed accordingly in a self-consistent way, since the saturation radius is given by requiring the continuity of the profile, namely, $\rho_{\mathrm{sat}} = \rho (r_{\mathrm{sat}})$:
\begin{equation}
\label{rsat}
r_{\mathrm{sat}} = R_{\mathrm{spike}} \left[ \dfrac{\rho_{\odot}}{\rho_{\mathrm{sat}}} \dfrac{r_{\odot}}{R_{\mathrm{spike}}} \left( 1 + \dfrac{r_{\odot}}{r_{\mathrm{s}}} \right) ^{2} \right] ^{1/\gamma_{\mathrm{spike}}}.
\end{equation}
Taking $ m_{\mathrm{DM}} = 10\ \rm GeV $, and assuming the canonical value of the cross section $ \left\langle \sigma v \right\rangle = 3 \times 10^{-26}\ \rm cm^{3}\ s^{-1} $, the saturation density given by annihilations is $ \rho_{\mathrm{sat}}^{\rm{ann}} \approx 1.06 \times 10^{9}\ \rm GeV\ cm^{-3} $. This leads to $ r_{\mathrm{sat}}^{\mathrm{ann}} \approx 5.3 \times 10^{-3}\ \rm pc $ for $ R_{\mathrm{spike}} = 1\ \rm pc $, $ \gamma_{\mathrm{spike}} = 7/3 $, and a conservative value of the infall time. For the NFW profile without a spike, the saturation radius is much smaller: $ r_{\mathrm{sat}}^{\mathrm{ann}} \approx 4.88 \times 10^{-6}\ \rm pc $. 

By combining Eqs.~\eqref{rhosat} and \eqref{rsat}, we see that $ r_{\mathrm{sat}}^{\mathrm{ann}} \propto \left\langle \sigma v \right\rangle^{1/\gamma_{\rm{spike}}} $ for the spike and $ r_{\mathrm{sat}}^{\mathrm{ann}} \propto \left\langle \sigma v \right\rangle $ for the NFW profile. Considering that $ \gamma_{\rm{spike}} > 1 $, the saturation radius is therefore much less dependent on $ \left\langle \sigma v \right\rangle $ for the spike than for NFW. We will also consider in the next sections a NFW$+$spike profile with a much smaller saturation radius, which is independent of the annihilation cross section. Typically we will choose $ r_{\mathrm{sat}} =r_{\mathrm{Sch}} = 4.2 \times 10^{-7}\ \rm pc $ with $r_{\mathrm{Sch}} $ the Schwarzschild radius of Sgr A*, leading to a saturation density of the order of $ 10^{18}\ \rm GeV\ cm^{-3} $. This is an extreme case that could correspond for instance to a very small infall time of DM particles onto the black hole.

\subsection{Loss term}
\label{losses}

For the propagation model to be complete, one must now specify the energy-loss term $ b( \vec{x},E) $. Here we neglect its spatial dependence and assume that the magnetic field is homogeneous over the entire diffusion zone. 

For the region of interest in this study, the dominant processes through which high energy electrons lose energy are synchrotron radiation and inverse Compton (IC) scattering on photons of the interstellar radiation field (ISRF). Bremsstrahlung losses are subdominant but we include them in the calculation. Coulomb losses are even smaller, but we include them for completeness. For both losses we use the expressions of Ref.~\cite{Sarazin}, with the electron density taken to be $ 1 \ \rm cm^{-3} $ \cite{BEnssS2}. Ionization losses are negligible for energies greater than $ 1\ \rm MeV $, and this condition is fulfilled for electrons produced in annihilations of $ 10\ \rm GeV $ DM particles, so we neglect them. The synchrotron energy-loss term is easy to quantify and reads \cite{Longair}
\begin{equation}
b_{\mathrm{syn}} = \dfrac{4}{3} \sigma_{\mathrm{T}} c \dfrac{B^{2}}{2 \mu_{0}} \gamma ^{2},
\end{equation}
where $ \sigma_{\mathrm{T}} $ is the Thomson cross section, $ B $ is the intensity of the magnetic field, $c$ is the speed of light, $\gamma$ is the Lorentz factor, and $ \mu_{0} $ is the vacuum permeability. 

\begin{table}[b]
\caption{\label{table} Temperatures and energy densities obtained by fitting the SED of the ISRF with greybody spectra. The parameters of the blackbody spectrum of the CMB are also displayed.}
\begin{ruledtabular}
\begin{tabular}{ccc}
 &$T\ \rm (K)$&$w \ \rm (GeV\ cm^{-3})$\\
\colrule
CMB & 2.725 & $2.602\times 10^{-10}$\\
IR & $4.231\times 10^{1}$ & $6.841\times 10^{-10}$\\
Stellar & $2.669\times 10^{2}$ & $1.214\times 10^{-10}$\\
 & $3.176\times 10^{3}$ & $3.317\times 10^{-9}$\\
UV & $6.373\times 10^{3}$ & $2.745\times 10^{-9}$\\
 & $2.437\times 10^{4}$ & $7.746\times 10^{-10}$\\
\end{tabular}
\end{ruledtabular}
\end{table}

Estimating the IC losses is more difficult. The reason is that IC losses can only be computed analytically for a blackbody distribution of photons. However, the ISRF does not follow a Planckian distribution since it is the sum of different components such as IR light from dust or optical and UV light from stars. The only true blackbody is the cosmic microwave background (CMB). We shall therefore follow the same procedure as in Ref.~\cite{Timur_ISRF}, except that we apply this method to the GC instead of the solar neighborhood. Such calculations are more precise than most calculations based on order of magnitude estimates of the synchrotron and IC characteristic times. However, one needs to recall that we have assumed that the losses are independent of the distance to the GC. This is only valid insofar as we focus on the inner region, where the synchrotron emission is expected to be dominant over synchrotron emission from regions more distant from the center. 

To apply the method of Ref.~\cite{Timur_ISRF}, we first use the spectral energy distribution (SED) of the ISRF given by the GALPROP team \cite{ISRF_Porter}. Considering that the electron propagation scale is smaller than $ 2\ \rm kpc $ in the energy range considered in this study, we average the SED on a cylinder of radius and half-height of $ 2\ \rm kpc $ centered on the GC, which is the region of interest in this paper. We then fit the averaged SED with greybody spectra characterized by the energy density $ w $ and the temperature $ T $. The SED of the ISRF is thus approximated by a sum of greybody spectra. The corresponding parameters of the fit are shown in Table~\ref{table}. The total energy-loss term for IC scattering is then the sum of the contributions of the IR, UV, stellar greybodies plus the CMB blackbody. The total energy-loss term $ b(E) $ is the sum of the loss terms for the IC, Bremsstrahlung, Coulomb, and synchrotron processes.

\subsection{Resolution of the transport equation: Halo function}

There exist several techniques in the literature to solve the transport equation. For instance, GALPROP relies on an implicit iteration scheme \cite{galprop} while USINE \cite{usine} and the method presented in Ref.~\cite{Itilda} are based on a semianalytical approach. Since GALPROP does not have the spatial resolution needed to zoom in on the GC (it has indeed a minimum step size of $ 10\ \rm pc $ due to the resolution of gas maps \cite{galprop}), we use the semianalytical method presented in Ref.~\cite{Itilda}.

\subsubsection{General features}

The main elements of the method of Ref.~\cite{Itilda} that we employ are summarized below. The spectrum $ \psi $ of the cosmic-ray particle after propagation is given by the expression
\begin{equation}
\psi (\vec{x},E) = \dfrac{\kappa}{b(E)} \int_{E}^{\infty} \! \tilde{I}_{\vec{x}}(\lambda_{\mathrm {D}}(E,E_{S})) \dfrac{\mathrm{d}n}{\mathrm{d}E}(E_{S}) \, \mathrm{d}E_{S},
\end{equation}
where $ \tilde{I}_{\vec{x}}(\lambda_{\mathrm {D}}(E,E_{S})) $ is called the halo function (computed in Sec.~\ref{Itilde_green}) and $ \kappa = (1/2) \left\langle \sigma v \right\rangle (\rho_{\odot}/m_{\mathrm{DM}})^{2} $ is defined by writing the source term as $ q = \kappa (\rho/\rho_{\odot})^{2} \mathrm{d}n/\mathrm{d}E $. The halo function encapsulates the information on propagation through the diffusion length $ \lambda_{\mathrm {D}} $. The latter represents the distance travelled by a particle produced at energy $ E_{S} $ and losing energy during propagation, down to energy $ E $. It is given by (see, e.g., Ref.~\cite{Synchrotron_wimps})
\begin{equation}
\lambda_{\mathrm {D}}^{2}(E,E_{S}) = 4 \int_{E}^{E_{S}} \! \dfrac{K(E')}{b(E')} \, \mathrm{d}E'.
\end{equation}
Under the assumption that the injection spectrum is a delta function, the flux after propagation takes on a simplified form:
\begin{equation}
\label{spectrum}
\psi (\vec{x},E) = \dfrac{\kappa}{b(E)} \tilde{I}_{\vec{x}}(\lambda_{\mathrm {D}}(E,m_{\mathrm{DM}})).
\end{equation}
In principle, the halo function $ \tilde{I} $ can then be computed using either a Fourier--Bessel series or a Green's function. In what follows we provide the reader with the expression of $ \tilde{I} $ in terms of a Green's function, but the expression in terms of Fourier--Bessel series can be found in Ref.~\cite{Itilda}. There exists actually another technique to compute $ \tilde{I} $ \cite{Zupan}. The latter consists in rewriting the transport equation as a partial differential equation for $ \tilde{I} $ and solving it numerically \cite{Zupan} or analytically \cite{Cirelli}. This third method is in principle fast and efficient since $ \tilde{I} $ can be computed only once for a given profile, but it is not convenient in our case because the radial dependence of spiky profiles leads to a stiff equation that requires an extremely large number of steps to be solved accurately, and there is no alternative in this case to reduce the computing time.

\subsubsection{Green's functions vs Fourier--Bessel series}

The Fourier--Bessel series approach relies on an expansion of the source term $ q $ as a series of cosines and Bessel functions \cite{Itilda}. The main element of the expansion is the factor $ j_{0} \left( \alpha_{i}r_{\mathrm{cyl}}/R_{\mathrm{gal}} \right) $, where $ j_{0} $ is the zeroth-order Bessel function of the first kind and $ \alpha_{i} $ the $i$th zero of $ j_{0} $ (with $ r_{\mathrm{cyl}} = \sqrt{x^{2} + y^{2}} $ in terms of Cartesian coordinates). The problem is that $ j_{0} \left( \alpha_{i}r_{\mathrm{cyl}}/R_{\mathrm{gal}} \right) $ goes to 1 when the argument (and therefore $ r_{\mathrm{cyl}} $) goes to 0, i.e.~toward the GC. The source term therefore appears to be constant while the DM halo profile continues to increase with small values of the radius. To prevent the argument of $ j_{0} $ from falling to zero too rapidly, one needs to sum over a large number of Bessel zeros $ \alpha_{i} $. So unless one uses a huge number of Bessel modes, the expansion cannot account for steep profiles on small scales, which leads to a halo function that is greatly underestimated at the center. However, taking, for example, $ 10^{9} $ modes results in an unacceptably long computing time. 

The Green's function approach allows us to avoid this difficulty, since we were able to define three different regimes for $ \lambda_{\mathrm {D}} $ (depending on the distance to the GC) to which the integration step can be adapted.

\subsubsection{Computing the halo function with Green's functions: General framework}
\label{Itilde_green}

Since the transport equation (\ref{transport equation}) is a diffusion equation, it can be rewritten as the heat equation in terms of a pseudotime related to the energy $ E $ via the diffusion length \cite{Green}. Consequently, the general solution can be expressed in terms of the propagator of the heat equation. However, one must take into account the boundaries of the diffusion zone, which leads to a different propagator from that corresponding to an infinite space. 

First, considering that the observer is located at $ d_{\mathrm{obs}} \equiv r_{\odot} = 8.5\ \rm kpc $ from the GC and since cosmic rays originate mostly from the central regions, it is safe to assume that the radial boundary at $ R_{\mathrm{gal}} = 20 \ \rm kpc $ has a negligible impact on the spectrum, especially for a medium half-thickness $ L $. Even for the half-thickness corresponding to the MAX set, the effect is small \citep{Itilda}. This infinite slab hypothesis allows one to write the propagator as the product of two independent factors corresponding to horizontal and vertical propagation \cite{Green}, \begin{align}
G(\vec{x},E \leftarrow \vec{x}_{\mathrm{S}},E_{\mathrm{S}}) = & \dfrac{1}{\lambda_{\mathrm {D}}^{2} \pi} \exp \left(- \dfrac{(x - x_{\mathrm{S}})^{2} + (y - y_{\mathrm{S}})^{2}}{\lambda_{\mathrm {D}}^{2}} \right) \nonumber \\
& \times V(z,E \leftarrow z_{\mathrm{S}},E_{\mathrm{S}}),
\end{align}
with $ \vec{x}_{\mathrm{S}} $ the position of production and $ \vec{x} $ the position after propagation. Here these positions are specified by Cartesian coordinates $ x_{\mathrm{S}} $, $ y_{\mathrm{S}} $, $ z_{\mathrm{S}} $, and $ x $, $ y $, $ z $ respectively. $ V $ is the vertical contribution to the propagator, for which different regimes arise. 

If the diffusion length of a cosmic ray is small enough, the particle does not feel the influence of the boundaries at $ z = \pm L $. Said more quantitatively, if $ \lambda_{\mathrm {D}}^{2} \ll L^{2} $ the free propagator is a very good approximation \cite{Green}:
\begin{equation}
V(z,E \leftarrow z_{\mathrm{S}},E_{\mathrm{S}}) = \dfrac{1}{\lambda_{\mathrm {D}} \sqrt{\pi}} \exp \left( - \dfrac{(z - z_{\mathrm{S}})^{2}}{\lambda_{\mathrm {D}}^{2}} \right).
\end{equation}

In the opposite regime, when $ \lambda_{\mathrm {D}}^{2} \gg L^{2} $, the propagation is sensitive to the vertical boundaries. As a result, the vertical propagator must be computed differently. However, the diffusion equation can be seen as a Schr\"odinger equation in imaginary time, so the diffusion equation can be interpreted as describing the evolution of a particle in the diffusion zone, which plays the part of an infinite potential well between $ z = -L $ and $ z = +L $. The vertical propagator may then be expanded as a series over the eigenfunctions of the associated Hamiltonian \cite{Green},
\begin{align}
\label{Vertical propagator}
V(z,E \leftarrow z_{\mathrm{S}},E_{\mathrm{S}}) = \dfrac{1}{L} \sum_{n=1}^{\infty} \left( \exp \left( -\dfrac{\lambda_{\mathrm {D}}^{2} k_{n}^{2}}{4} \right) \varphi_{n}(z_{\mathrm{S}}) \varphi_{n}(z) \right. \nonumber \\ 
\left. + \exp \left( -\dfrac{\lambda_{\mathrm {D}}^{2} k_{n}^{\prime 2}}{4} \right) \varphi'_{n}(z_{\mathrm{S}}) \varphi'_{n}(z) \right),
\end{align}
where the wave functions $ \varphi_{n} $ and $ \varphi'_{n} $ are, respectively, even and odd: $ \varphi_{n}(z) = \sin (k_{n}(L-|z|)) $ and $ \varphi'_{n}(z) = \sin (k'_{n}(L-z)) $, with the wave vectors defined as $ k_{n} = \left( n-1/2 \right) \pi/L$ and $ k'_{n} = n \pi/L $. When the diffusion length is large enough, the series in Eq.~\eqref{Vertical propagator} can be truncated to less than 100 terms. We have used $ 0.5\ \rm kpc $ as the limiting value between these two regimes.

Once the propagator $ G $ is known, the halo function is given by the convolution of $ G $ with the source term, namely, the DM density squared, over the diffusion zone (DZ) \cite{Itilda}:
\begin{equation}
\label{Itilde_convolution}
\tilde{I}_{\vec{x}}(\lambda_{\mathrm {D}}(E,E_{S})) = \int_{\mathrm{DZ}} \! \mathrm{d}\vec{x}_{\mathrm{S}} \, G(\vec{x},E \leftarrow \vec{x}_{\mathrm{S}},E_{\mathrm{S}}) \left( \dfrac{\rho(\vec{x}_{\mathrm{S}})}{\rho_{\odot}} \right) ^{2}.
\end{equation}
However, depending on the value of $ \lambda_{\mathrm {D}} $ relative to the distance from the GC, the propagator can become very sharply peaked. Moreover, the DM profile is also very sharply peaked. Consequently, if the sampling of the integrand is not done properly, the divergence is completely missed, and the halo function is underestimated. For the sharpness of the profile, we use logarithmic steps, but the sharpness of the propagator requires a more complex treatment detailed in the following.

\subsubsection{Computing the halo function with Green's functions: Trick for the propagator}

\begin{figure*}[tbp]
\centering 
\includegraphics[scale=0.41]{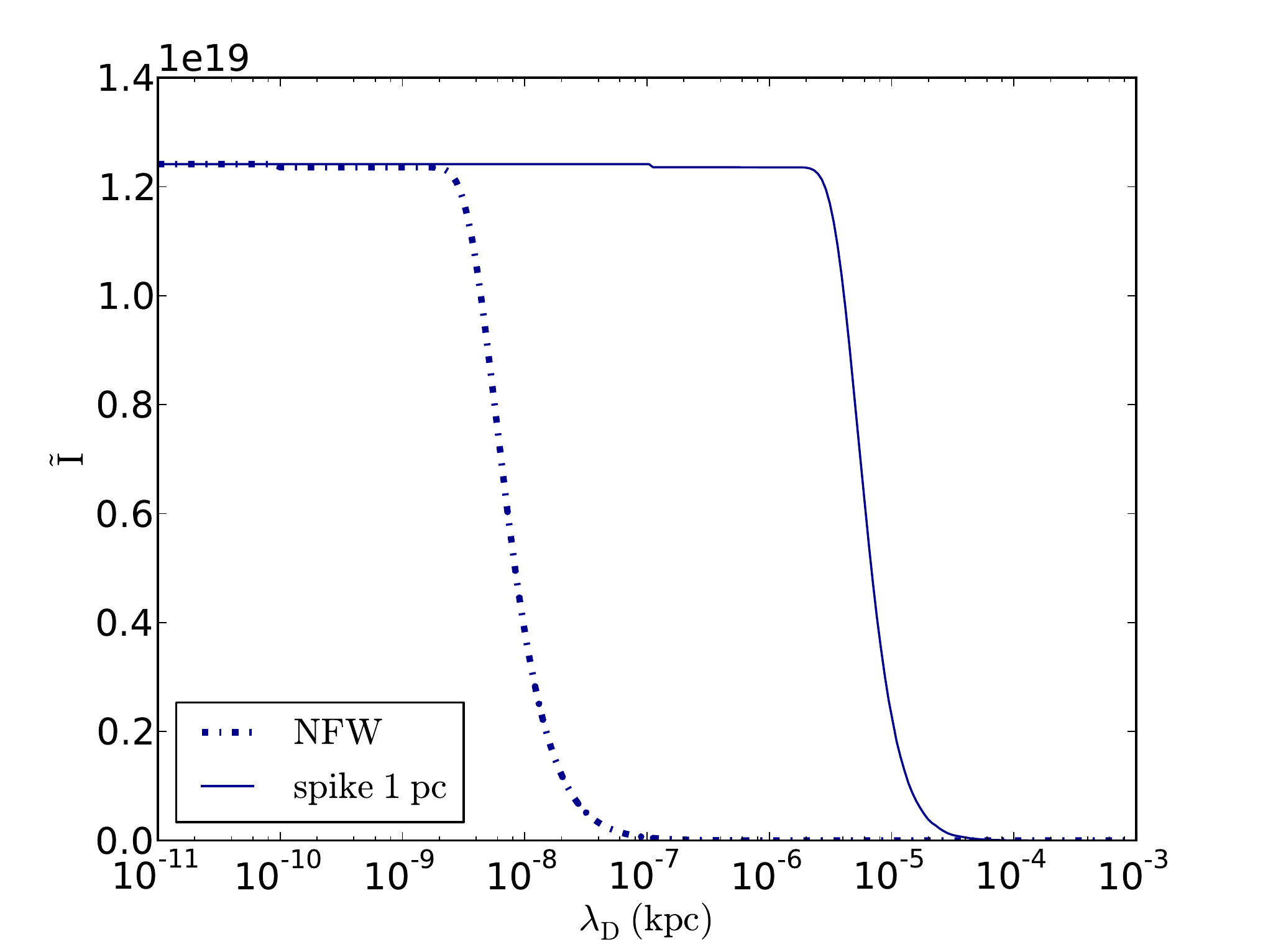}\includegraphics[scale=0.41]{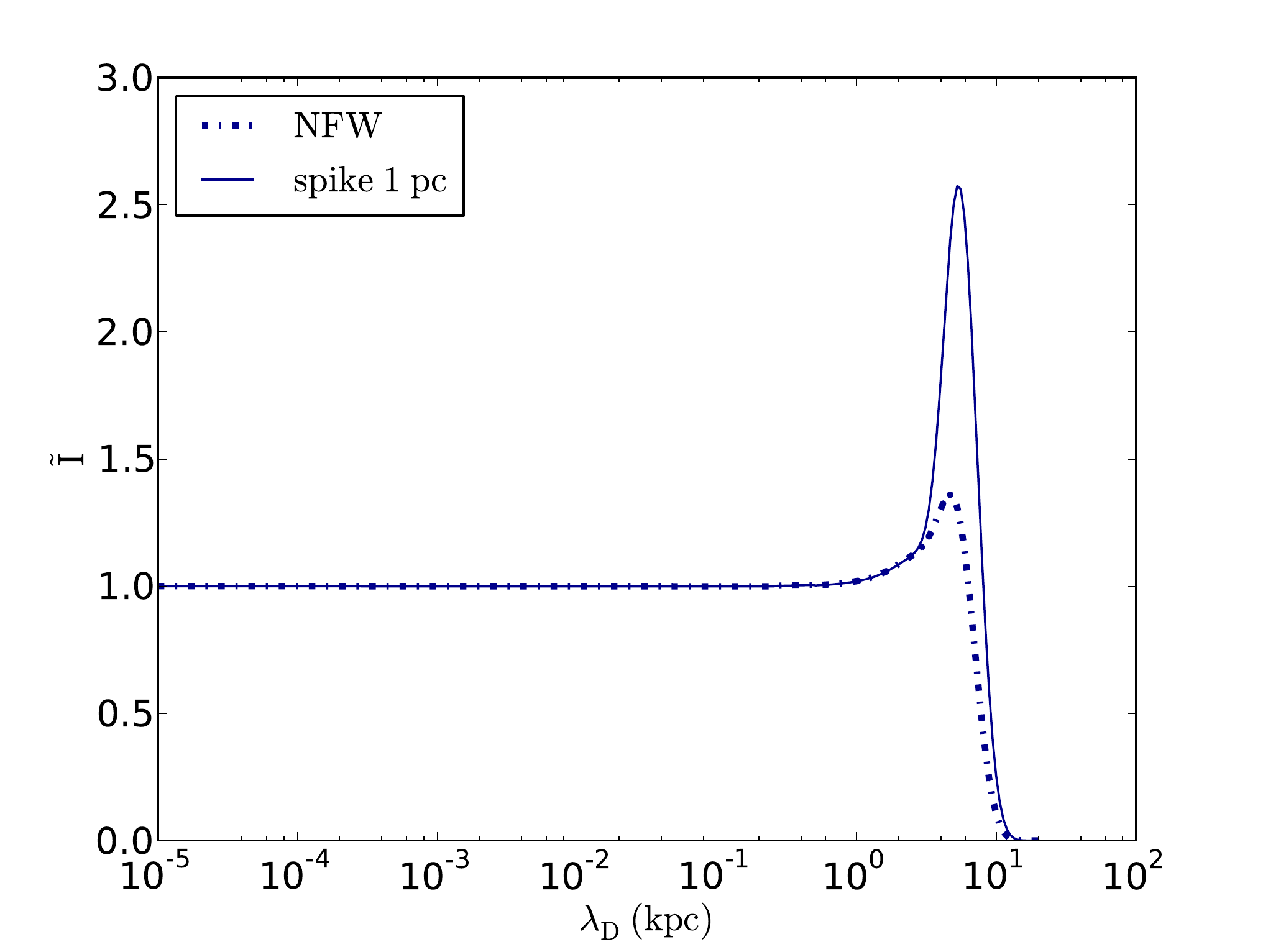} 
\caption{\label{Itilde} Halo function at the GC (left panel) and at the position of the Sun (right panel), as a function of the diffusion length, for the NFW profile (dashed-dotted line) and the NFW+spike profile with $ R_{\rm{spike}} = 1\ \rm pc $ (solid line). Here we use the MED parameter set.}
\end{figure*}

Our new method consists in computing the halo function at small scales by exploiting the three different regimes that arise for the horizontal and vertical components of the propagator, depending on the value of $ \lambda_{\mathrm {D}} $ relative to $ r_{\rm{cyl}} = \sqrt{x^{2} + y^{2}} $ and $ z $. 

First of all, in the regime of small $ \lambda_{\mathrm {D}} $, i.e., for $ \lambda_{\mathrm {D}} \ll r_{\rm{cyl}} $ or $ \lambda_{\mathrm {D}} \ll z$, the horizontal and vertical components of the propagator become extremely sharply peaked. In this case, a huge number of integration steps would be required to capture the peak in the integral. However, the halo function has an analytic limit for $ \lambda_{\mathrm {D}} $ going to zero. Indeed, for $ \lambda_{\mathrm {D}} \rightarrow 0$, the propagator $ G(\vec{x},E \leftarrow \vec{x}_{\mathrm{S}},E_{\mathrm{S}}) $ becomes a delta function of space, $ \delta (\vec{x} - \vec{x}_{\mathrm{S}})$. Consequently, taking the limit of Eq.~\eqref{Itilde_convolution} for $ \lambda_{\mathrm {D}} $ going to 0, or equivalently $ E $ going to $ E_{\mathrm{S}} $, leads to
\begin{equation}
\label{Itilde_limit}
\tilde{I}_{\vec{x}}(\lambda_{\mathrm {D}}) \underset{\lambda_{\mathrm {D}}\to 0}{\longrightarrow} \left( \dfrac{\rho(\vec{x})}{\rho_{\odot}} \right) ^{2},
\end{equation}
which is equal to 1 at the Sun's position ($ \tilde{I}_{\odot} = 1 $) and very large (depending on the type of spike that we consider) at the GC. Therefore, to solve the problem of the sharply peaked propagator missed by the integral for $ \lambda_{\mathrm {D}} \ll r_{\rm{cyl}} $ or $ \lambda_{\mathrm {D}} \ll z$, we have imposed by hand the condition displayed in Eq.\eqref{Itilde_limit} in this regime. This way we ensure that the value of $ \tilde{I} $ is correct when cosmic rays do not propagate.

In the intermediate regime, when the propagators are peaked but with finite widths, we compute the spatial integrals over such widths instead of integrating over the whole range of values of $ r_{\mathrm{S}} $ or $ z_{\mathrm{S}} $. This is essential since the analytic limit is no longer a good approximation in this regime, and unless one uses a huge number of points, the integration procedure over the whole range once again misses the peak. Finally, for larger values of $ \lambda_{\mathrm {D}} $, i.e.\ when $ \lambda_{\mathrm {D}} \sim r_{\rm{cyl}} $ or $ \lambda_{\mathrm {D}} \sim z $, $ \tilde{I} $ is computed by doing the complete integrals over the diffusion zone. 

Using this adaptive procedure enables us to derive the halo function at the GC. Shown in Fig.~\ref{Itilde} (left panel) are the corresponding curves for the NFW profile and the NFW+spike profile, where we assume $ R_{\mathrm{spike}} = 1\ \rm pc $, $ r_{\mathrm{sat}} = r_{\mathrm{sat}}^{\mathrm{ann}} $, $ \left\langle \sigma v \right\rangle = 3 \times 10^{-26}\ \rm cm^{3}\ s^{-1} $, $ m_{\mathrm{DM}} = 10\ \rm GeV $, and the MED parameter set given in Eq.~\eqref{propa_sets}. As can be seen in this figure, the reconstruction works well, since the numerical solution reaches the plateau corresponding to the analytical solution when $\lambda_{\mathrm {D}} \rightarrow 0$. The relative error between the numerical and analytical solutions is smaller than the percent level, as shown by the small step at roughly $ 10^{-7}\ \rm kpc $. Note that we obtain similar results for a spike with $ r_{\mathrm{sat}} = r_{\mathrm{Sch}} $. 

In the right panel of Fig.~\ref{Itilde}, we also reproduce the halo function at the Sun's position ($ \tilde{I}_{\odot} $) as a function of $ \lambda_{\mathrm {D}} $ for the NFW profile (see Ref.~\cite{Itilda}) with $ r_{\mathrm{sat}} = r_{\mathrm{sat}}^{\mathrm{ann}} $ and a NFW$+$spike profile with $ R_{\mathrm{spike}} = 1\ \rm pc $ and $ r_{\mathrm{sat}} = r_{\mathrm{sat}}^{\mathrm{ann}} $. In this plot we have assumed the MED parameter set, $ \left\langle \sigma v \right\rangle = 3 \times 10^{-26}\ \rm cm^{3}\ s^{-1} $ and $ m_{\mathrm{DM}} = 10\ \rm GeV$. 
 
Armed with this very precise treatment of the halo function (and the resulting spectrum of primary electrons and positrons after propagation) for very small $ \lambda_{\mathrm {D}} $ and very small distances from the GC, we can now estimate the synchrotron flux from DM annihilations.

\subsection{Synchrotron flux}
\label{synchrotron}

\begin{figure}[h!]
\centering
\includegraphics[scale=0.25]{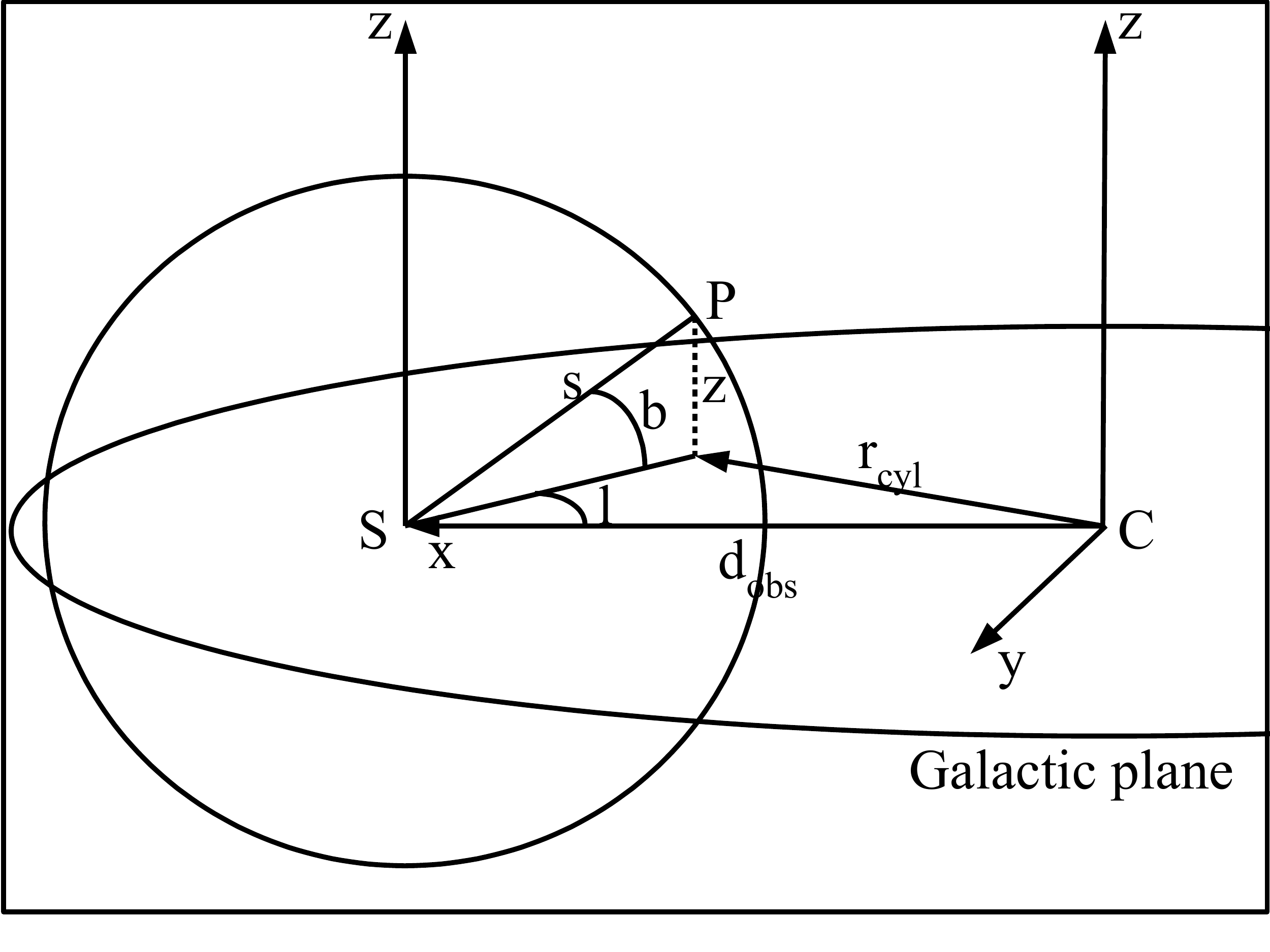} 
\caption{Coordinate systems for cosmic rays in the Galaxy. For propagation, cylindrical coordinates centered on the GC denoted as C are used. Sky maps are based on spherical coordinates centered on the Sun S. $ l $ and $ b $ are the longitude and latitude of the observed point P, and $ s $ is the radial coordinate along the line of sight.}
\label{coordinate_systems}
\end{figure}

The synchrotron power per unit frequency reads (see, e.g., \cite{Longair})
\begin{equation}
P_{\mathrm{syn}}(E,\nu) = \dfrac{1}{4 \pi \epsilon_{0}} \dfrac{\sqrt{3}e^{3}B}{m_{\mathrm{e}}c} F_{\rm{i}} \left( \dfrac{\nu}{\nu_{\mathrm{c}}} \right),
\end{equation}
where $ m_{\mathrm{e}}$ is the electron mass, $ e $ the elementary charge, $ \epsilon_{0} $ the vacuum permittivity, and the critical frequency is given by
\begin{equation}
\nu_{\mathrm{c}} = \dfrac{3eE^{2}B}{4 \pi m_{\mathrm{e}}^{3}c^{4}}.
\end{equation}
$ F_{\rm{i}} $ is the isotropic synchrotron spectrum, which accounts for the isotropic propagation of cosmic rays. In Ref.~\cite{Synchrotron_wimps}, the authors have shown that this function can be fitted by
\begin{equation}
F_{\rm{i}}(x) = a x^{d} \exp \left( -\sqrt{\dfrac{x}{b}} - \dfrac{x}{c} \right), 
\end{equation}
where $ x = \nu / \nu_{\mathrm{c}} $ and the four parameters of the best fit are $ a = 1.60883 $, $ b = 1.95886 $, $ c = 1.13147 $, and $ d = 0.33839 $. We use this parametrization in this work. From there, the synchrotron emissivity reads (see Ref.~\cite{Synchrotron_wimps})
\begin{equation}
j_{\nu}(\vec{x}) = N_{\mathrm{e}} \int_{m_{\mathrm{e}}}^{m_{\mathrm{DM}}} \! P_{\mathrm{syn}}(E,\nu) \psi _{\mathrm{e}}(\vec{x},E) \, \mathrm{d}E, 
\end{equation}
where $ \psi _{\mathrm{e}} $ is the electron spectrum after propagation and $ N_{\mathrm{e}} = 2 $. For making maps of the synchrotron emission, we will use the relations between the Cartesian coordinates and longitude $ l $ and latitude $ b $ obtained by considering the geometry of the diffusion zone (see Fig.~\ref{coordinate_systems}, Appendix), namely,
\begin{equation}
\label{xyz}
x = d_{\mathrm{obs}} - s \cos b \cos l,\ y = -s \cos b \sin l,\ z = s \sin b,
\end{equation}
with $ s $ the radial coordinate along the line of sight (l.o.s.\!). Finally, the synchrotron flux received at the Earth from the direction $ (l,b) $ is derived by integrating the emissivity $ j_{\nu}(\vec{x}) \equiv j_{\nu}(s,l,b) $ at frequency $ \nu $ over $ s $ in the direction defined by $ l $ and $ b $ \cite{Synchrotron_DM_decay}:
\begin{equation}
\Phi_{\nu}(l,b) = \dfrac{1}{4\pi} \int_{\mathrm{l.o.s.}} \! j_{\nu}(s,l,b) \, \mathrm{d}s.
\end{equation}

\section{Constraining the existence of a dark matter spike}
\label{results}

\begin{figure*}[tbp]
\centering
\includegraphics[scale=0.3]{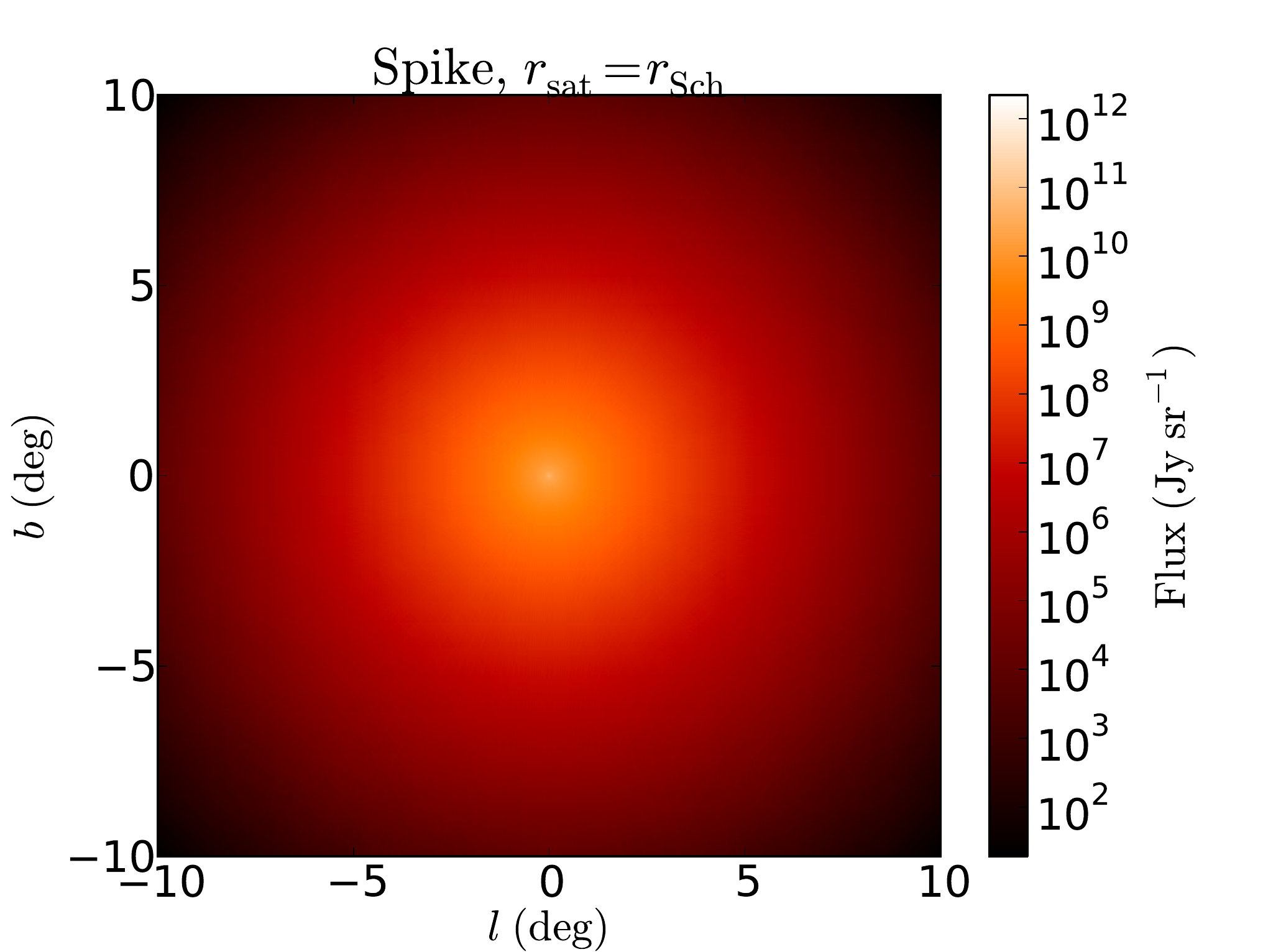}\includegraphics[scale=0.3]{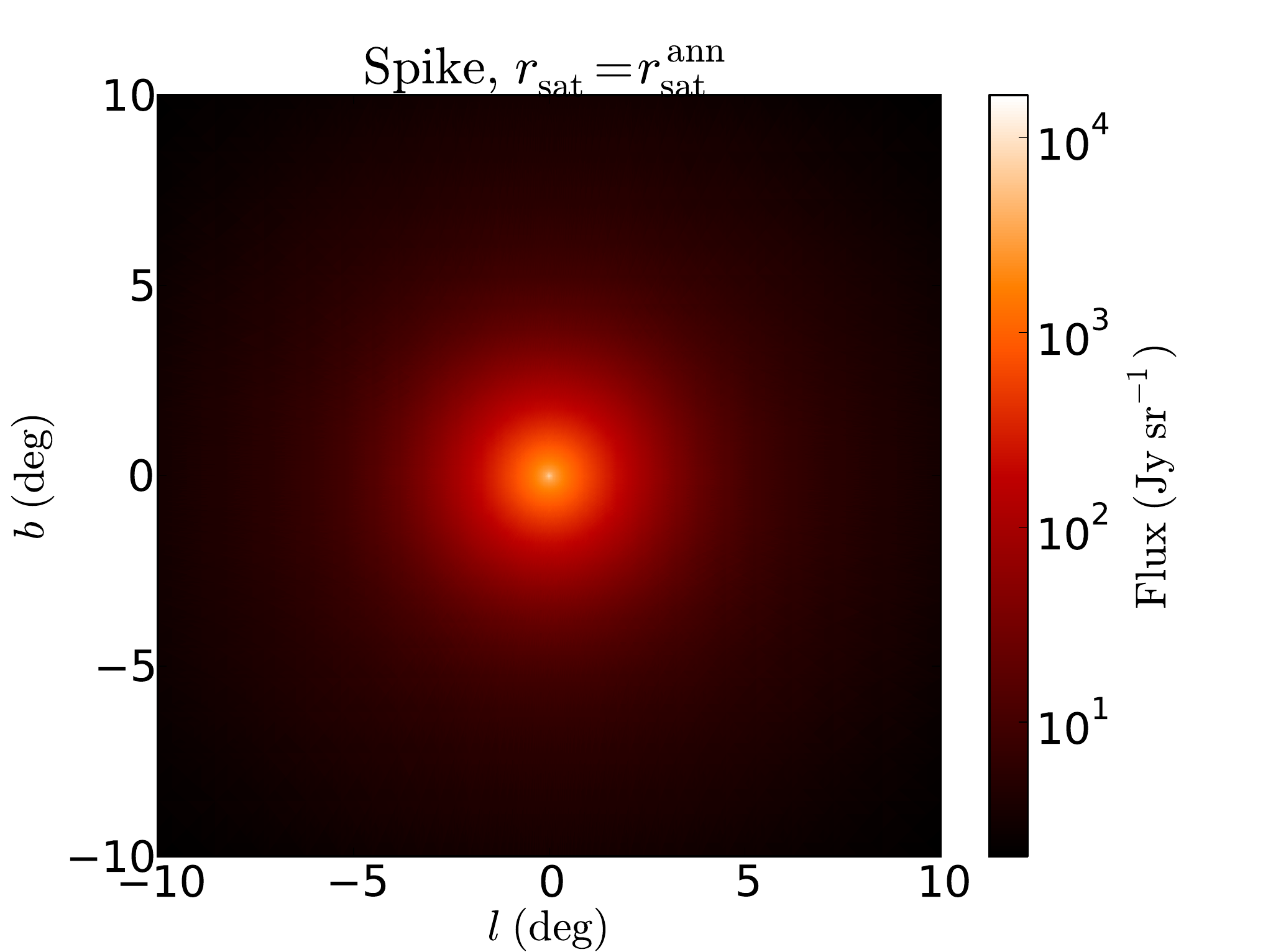}

\includegraphics[scale=0.3]{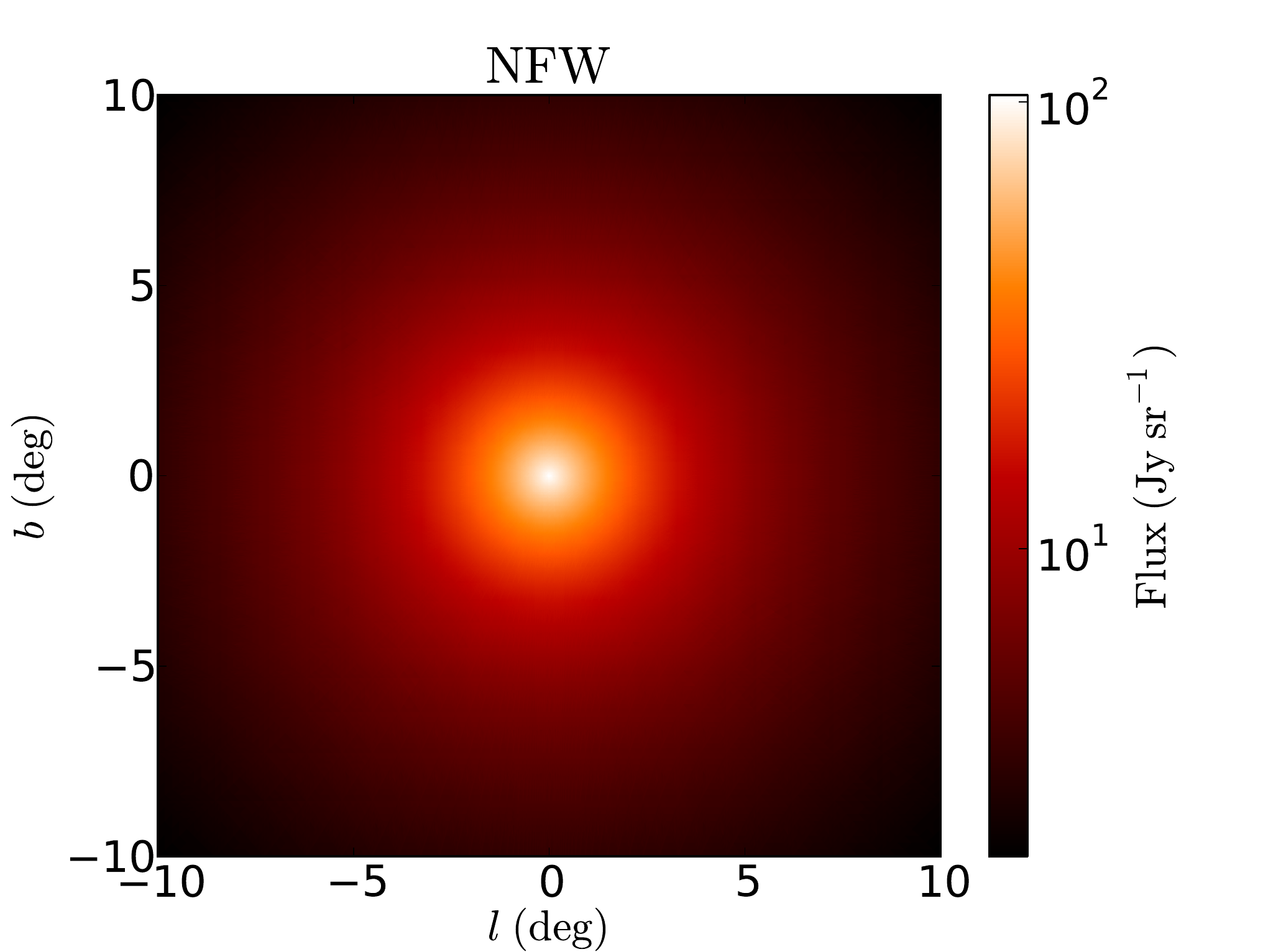} 
\caption{\label{maps}$ 30 \ \rm GHz $ maps of the synchrotron flux induced by $ 10 \ \rm GeV $ DM particles, for $ \left\langle \sigma v \right\rangle = 3 \times 10^{-26}\ \rm cm^{3}\ s^{-1} $, $ B = 3\ \rm \mu G $, and the MED set of propagation parameters. The DM profiles used are spikes with $ \gamma_{\mathrm{spike}} = 7/3 $, $ R_{\mathrm{spike}} = 1 \ \rm pc $, with $ r_{\mathrm{sat}} = r_{\mathrm{Sch}} $ (left panel), $ r_{\mathrm{sat}} = r_{\mathrm{sat}}^{\mathrm{ann}} $ (middle panel), and the NFW profile (right panel). For the spike with $ r_{\mathrm{sat}} = r_{\mathrm{Sch}} $, the flux varies by 10 orders of magnitude between the inner region (a few $ \mu \rm as $) and $ 10^{\circ} $ from the center.}
\end{figure*}

With our new technique for the treatment of cosmic-ray propagation in the inner Galaxy, we can now attempt to determine whether it is possible to distinguish a spiky DM halo profile from a NFW distribution and whether one can constrain the properties of the spike using synchrotron emission. In the next sections, we will mostly consider light DM particles (typically $ m_{\mathrm{DM}} = 10 \ \rm GeV $), but we will show that our conclusions remain valid in the case of heavy DM particles.

\subsection{Morphology of the synchrotron emission: Maps of the GC with or without a spike}
\label{morpho_maps}

The presence of a spike in the dark matter halo profile is expected to affect the morphology of the synchrotron emission coming from DM particles. The latter can be inferred by looking at synchrotron maps in terms of longitude $l$ and latitude $b$ \cite{Planck_DelahayeBoehmSilk}. For a $ 10\ \rm GeV $ WIMP and relatively low values of the magnetic field, one expects a signal in the lowest frequency channels of the Planck low frequency instrument (LFI), in particular at $ 30\ \rm GHz $, and no other signature in any of the Planck high frequency instrument (HFI) channels. 

To establish these maps, we use the canonical value of $ 3 \times 10^{-26}\ \rm cm^3\ s^{-1} $ for the annihilation cross section, a constant value of $ 3 \ \rm \mu G $ for the magnetic field $ B $, and the MED set of diffusion parameters unless stated otherwise. The results are shown in Fig.~\ref{maps}. The left panel shows the synchrotron emission in the extreme case of a NFW+spike profile with $ R_{\mathrm{spike}} = 1\ \rm pc $ and $ r_{\mathrm{sat}} = r_{\mathrm{Sch}} = 4.2 \times 10^{-7}\ \rm pc $. A more realistic case, corresponding to a NFW+spike profile with $ R_{\mathrm{spike}} = 1\ \rm pc $ and $ r_{\mathrm{sat}} = r_{\mathrm{sat}}^{\mathrm{ann}} \approx 5.3 \times 10^{-3}\ \rm pc $, is displayed in the middle panel, while the NFW case is shown in the right panel. 

By comparing the left and middle panels, we see that the spike with the smallest saturation radius ($ r_{\mathrm{sat}} = r_{\mathrm{Sch}} $) leads to an extremely bright synchrotron emission (very high flux) very close to the GC. This is due to the very large number density of electrons injected in the center and an inefficient diffusion, as explained in Sec.~\ref{small scales}. We see also that the emission in the case of a NFW profile (right panel) is much more extended than for spiky profiles for a similar reason: the density is much lower on larger scales, and diffusion is more efficient. Hence, different DM halo profiles predict distinctive morphological signatures and synchrotron fluxes. Therefore, the combination of both the normalization and the morphology of the flux could be used to probe the existence of a spike in the inner Galaxy. 

This conclusion is in agreement with that from Ref.~\cite{ascasibar,BoehmDelahayeSilk}, where the morphology was used to distinguish decaying from annihilating DM scenarios (i.e., $\rho$ vs $\rho^2$). But more importantly, these maps also indicate that very steep profiles in the GC have signatures visible on scales of a few degrees (i.e., at much larger scales than $R_{\mathrm{spike}}$).

As a result one may be able to distinguish the DM energy distribution in the very inner Galaxy, even in the absence of synchrotron measurements at these scales. This new and very important result already suggests that even the Planck data may have the potential to constrain spiky profiles.

\subsection{Can we distinguish different inner profiles using their synchrotron emission?}

Maps are well suited for highlighting the morphology of the signal, but not for quantitatively comparing the fluxes associated with different profiles. Therefore, we now study the dependence of the synchrotron flux in terms of latitude. In practice, one should investigate the dependence in terms of $l$ and $b$, but giving the results in terms of latitude is actually enough, given the symmetry of the source (the latitude being slightly more relevant as the effects of the diffusion zone are more noticeable in this direction).

\subsubsection{Large scales (a few degrees)}

\begin{figure}[t]
\centering
\includegraphics[scale=0.41]{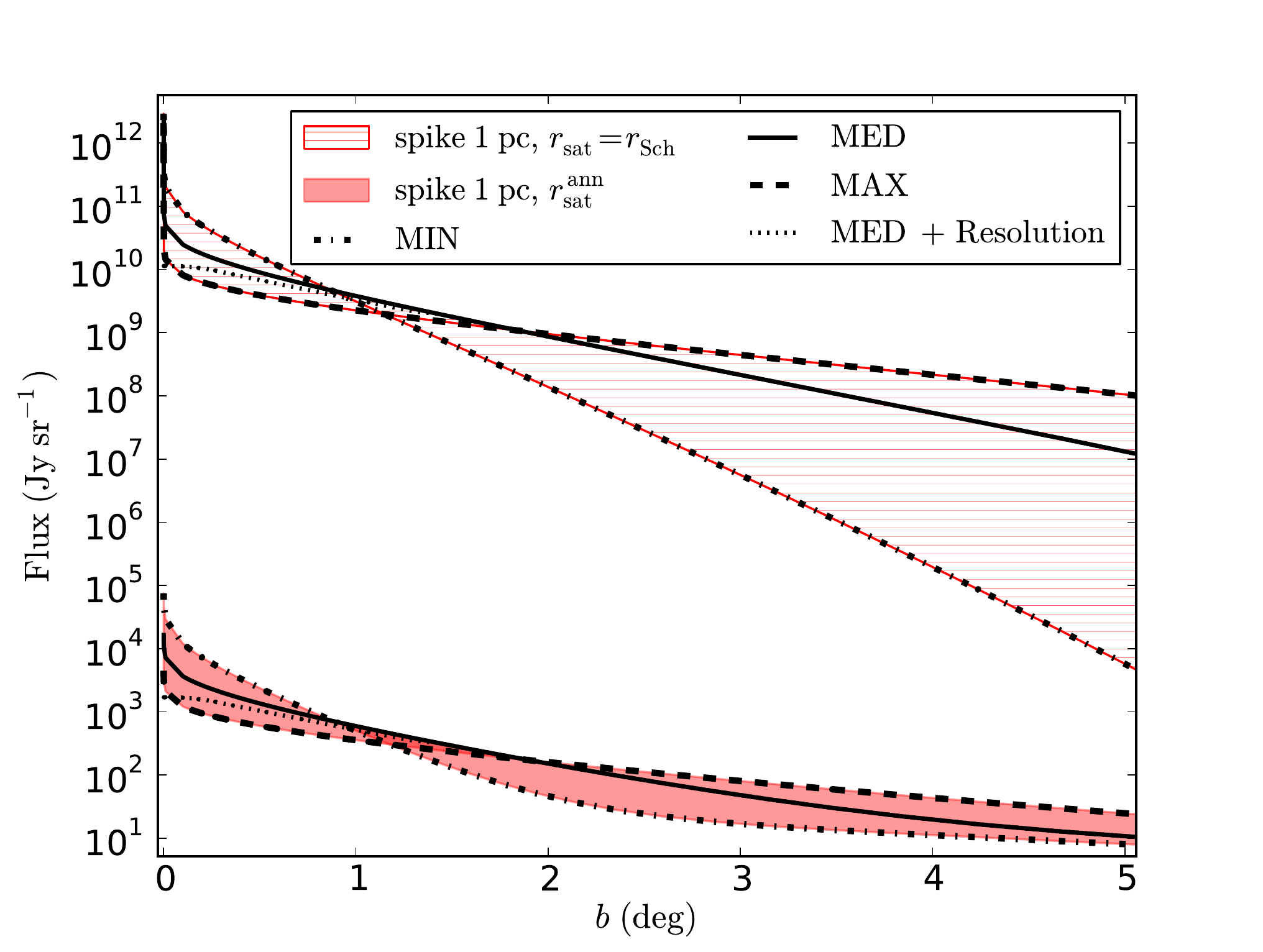} 
\caption{\label{flux_30GHz_10GeV_3microG_3e-26_latitude}Synchrotron flux as a function of latitude $ b $, for $ 10 \ \rm GeV $ DM particles, $ \left\langle \sigma v \right\rangle = 3 \times 10^{-26}\ \rm cm^{3}\ s^{-1} $, $ B = 3\ \rm \mu G $ and $ \nu = 30\ \rm GHz $. The red horizontally hatched and shaded areas represent the flux for a spike with $ \gamma_{\mathrm{spike}} = 7/3 $ and $ R_{\mathrm{spike}} = 1\ \rm pc $, respectively, for $ r_{\mathrm{sat}} = r_{\mathrm{Sch}} $ and $ r_{\mathrm{sat}} = r_{\mathrm{sat}}^{\mathrm{ann}} $. The uncertainty on the diffusion model is defined by the flux for the MIN (dashed-dotted lines) and MAX (dashed lines) propagation parameters. The solid lines are associated to the MED set. The dotted lines represent the flux for the MED set smoothed using the angular resolution of LFI at $ 30\ \rm GHz $, namely, $ 33 \ \rm arcmin $.}
\end{figure}

To begin with, we shall consider relatively large scales ($0.1^{\circ} \lesssim b \lesssim 10^{\circ} $). Our synchrotron predictions for those scales are shown in Fig.~\ref{flux_30GHz_10GeV_3microG_3e-26_latitude}, still assuming $ \left\langle \sigma v \right\rangle = 3 \times 10^{-26}\ \rm cm^{3}\ s^{-1} $ and $ B = 3\ \rm \mu G $.

Since one should in principle take into account the resolution of the detector, we first compute the average of the flux over the solid angle $ \Delta \Omega \approx \pi \theta_{\mathrm{res}}^{2} $, where $ \theta_{\mathrm{res}} $ is the resolution of the instrument, namely, $ 33 \ \rm arcmin $ at $ 30\ \rm GHz $ for Planck/LFI \cite{LFI_beams}:
\begin{equation}
\left\langle \Phi _{\nu} (l,b) \right\rangle _{\Delta \Omega} = \dfrac{1}{\Delta \Omega} \int_{\Delta \Omega} \! \Phi _{\nu} (l',b') \, \mathrm{d}\Omega',
\end{equation}
The corresponding result is shown as dotted lines in Fig.~\ref{flux_30GHz_10GeV_3microG_3e-26_latitude} (visible below $ 1^{\circ}$) in the case of a spiky profile with $R_{\mathrm{spike}} = 1\ \rm pc $, the MED set of parameters, and $ r_{\mathrm{sat}}$ equal to either $ r_{\mathrm{Sch}} $ or $r_{\mathrm{sat}}^{\mathrm{ann}} $. 

In both cases, accounting for the angular resolution of Planck at $ 30\ \rm GHz $ reduces the flux in the inner region by only less than 1 order of magnitude (making the emission look more extended). Since this does not have a significant impact on the estimates of the flux and adding an extra integral slows down our calculations, we do not average over the resolution of the detector. This also allows us to keep our results independent of a particular experiment.

Figure \ref{flux_30GHz_10GeV_3microG_3e-26_latitude} also enables us to study the impact of the saturation radius. For $ R_{\mathrm{spike}} =1\ \mathrm{pc}$, we can compare the synchrotron fluxes as a function of latitude for $ r_{\mathrm{sat}} = r_{\mathrm{sat}}^{\mathrm{ann}} $ and $ r_{\mathrm{sat}} = r_{\mathrm{Sch}} $. The spike with the extremely small saturation radius ($ r_{\mathrm{sat}} = r_{\mathrm{Sch}} $) predicts a flux that is orders of magnitude greater than that for the spike with $ r_{\mathrm{sat}}^{\mathrm{ann}} $. This is true both at $0.1^\circ$ and $10^\circ$. Since the value of $r_{\mathrm{sat}}$ affects the normalization of the flux on visible scales, it should be possible to distinguish spikes which have the same size but different saturation radii by measuring the synchrotron flux at latitude $b \sim {\cal{O}}(1^\circ)$. This is consistent with the preliminary conclusion obtained in Sec.~\ref{morpho_maps}, using the synchrotron maps. 

\begin{figure*}[tbp]
\centering 
\includegraphics[scale=0.41]{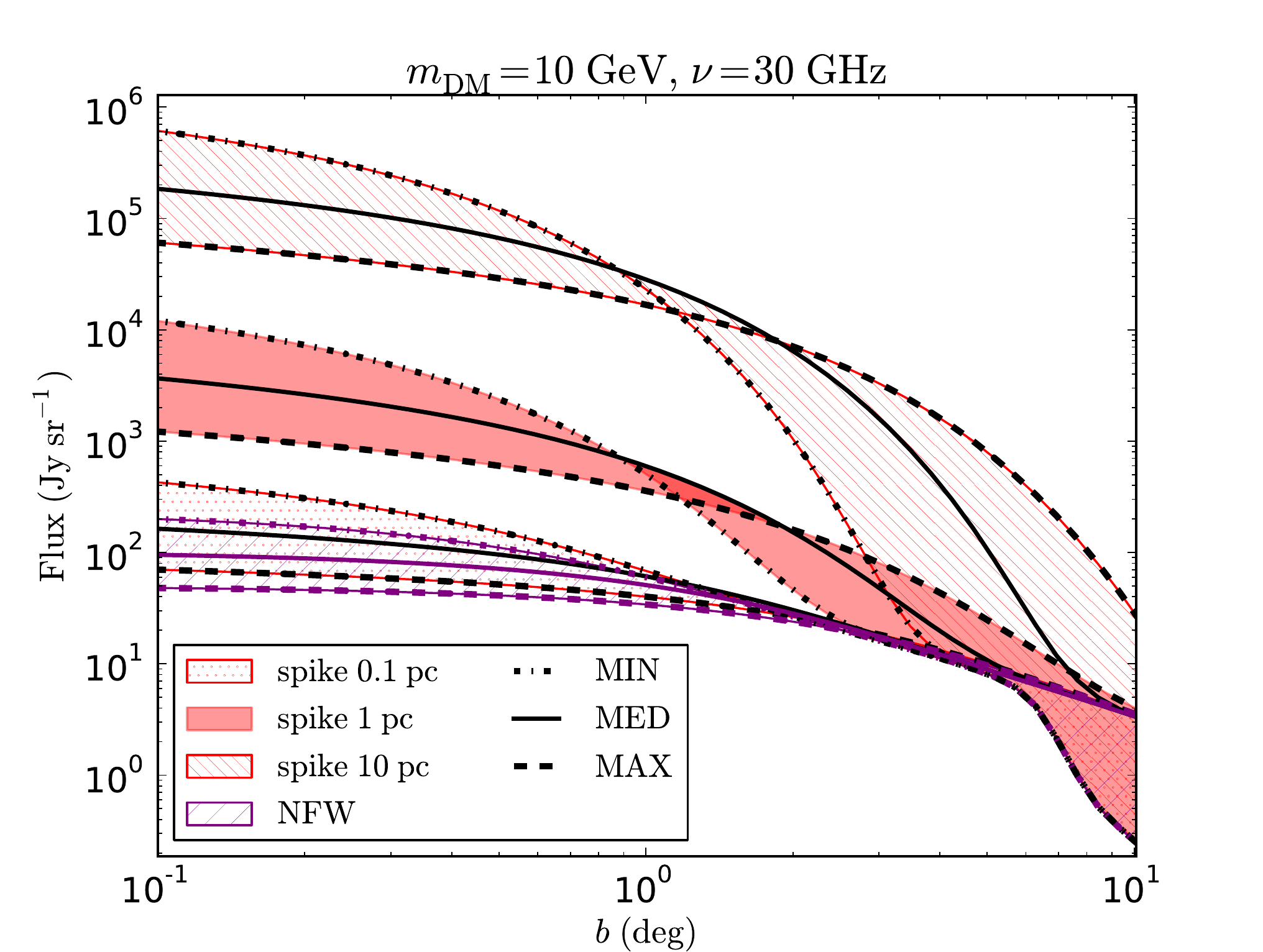}\includegraphics[scale=0.41]{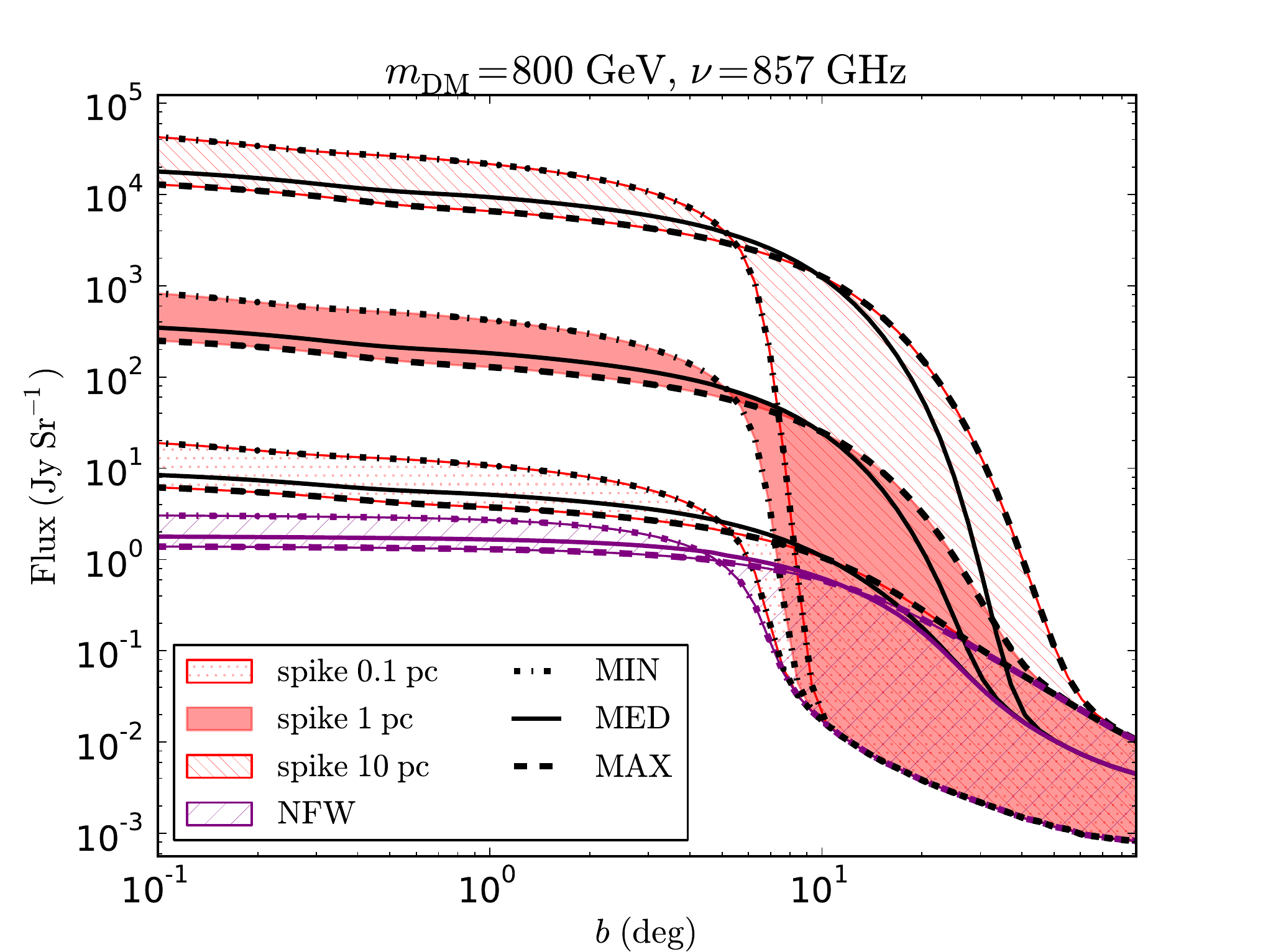}
\caption{\label{flux 3 spikes}Synchrotron flux as a function of latitude $ b $, for $ \left\langle \sigma v \right\rangle = 3 \times 10^{-26}\ \rm cm^{3}\ s^{-1} $, and $ B = 3\ \rm \mu G $. The spikes are characterized by $ \gamma_{\mathrm{spike}} = 7/3 $, $ r_{\mathrm{sat}} = r_{\mathrm{sat}}^{\mathrm{ann}} $, and different radii. The left panel corresponds to $ m_{\mathrm{DM}} = 10 \ \rm GeV $ and $ \nu = 30\ \rm GHz $, while the right panel corresponds to $ m_{\mathrm{DM}} = 800 \ \rm GeV $ and $ \nu = 857 \ \rm GHz $.. The red dotted, shaded, and hatched areas represent the flux for a spike of radius $ 0.1 $, $ 1 $ and $ 10\ \rm pc $ respectively. The purple hatched area is the flux for the NFW profile without a spike.}
\end{figure*}

Let us now study how the size of the spike $ R_{\mathrm{spike}} $ affects the flux. For this purpose, we fix $ r_{\mathrm{sat}}$ to $r_{\mathrm{sat}}^{\mathrm{ann}}$. Figure \ref{flux 3 spikes} (left panel) shows that different values of $ R_{\mathrm{spike}} $ induce a distinctive morphology: fluxes indeed decrease differently with latitude depending on $ R_{\mathrm{spike}} $. The only exception is when $ R_{\mathrm{spike}} = 0.1\ \rm pc $ as the morphology of the flux in this case is somewhat degenerated with the predictions for a NFW profile. For all the other profiles, it should be possible to determine the size of the spike $ R_{\mathrm{spike}} $ by looking at the synchrotron flux around $b \sim 1^{\circ} $. 

Therefore, one can constrain both the existence of a spike in the DM density at the GC and its size using both the normalization of the flux of the synchrotron emission and its morphology at degree scales. 

Our conclusions are similar in the case of heavy DM (see Fig.~\ref{flux 3 spikes}, right panel). In this figure we show the synchrotron flux for $ 800\ \rm GeV $ DM particles and a frequency of $ \nu = 857\ \rm GHz $ (the highest frequency channel of Planck/HFI). As one can see, spiky profiles with spikes of different sizes lead to a different morphology of the flux below $ 10^{\circ} $. The main uncertainty on the value of the flux actually arises from diffusion, since at such energies electrons diffuse more toward outer regions of the Galaxy and are thus more sensitive to the boundaries of the diffusion zone. However, keeping this caveat in mind, the morphology of the synchrotron emission can also be used to constrain the existence of a spike and its characteristics if DM is made of heavy particles.

\subsubsection{Zooming in on the very center (subarcsecond scales)}
\label{small scales}

\begin{figure*}[tbp]
\centering 
\includegraphics[scale=0.41]{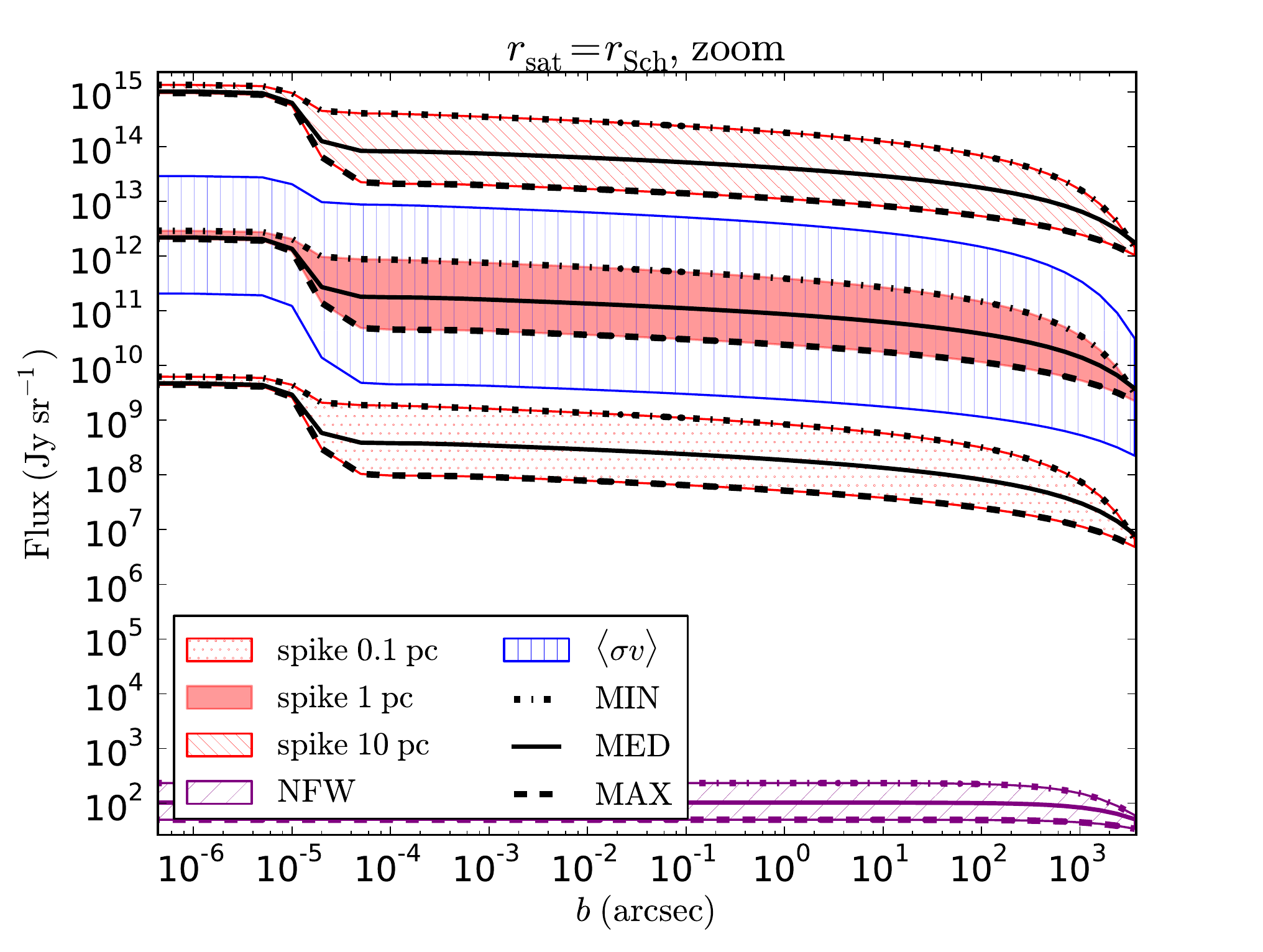}\includegraphics[scale=0.41]{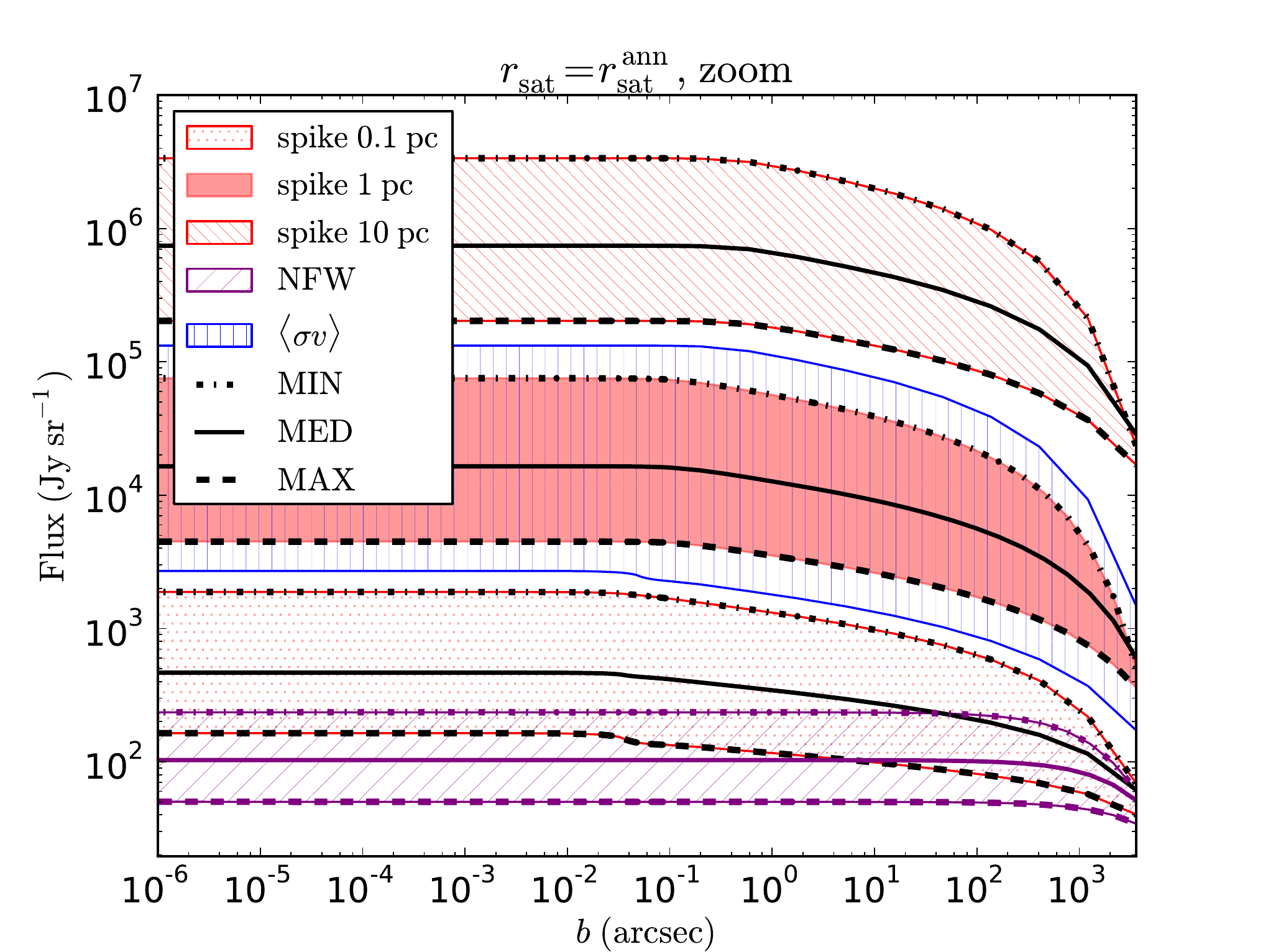}
\caption{\label{flux spike zoom}Synchrotron flux from the inner part of the Galaxy as a function of latitude $ b $, for $ 10\ \rm GeV $ DM particles, $ \left\langle \sigma v \right\rangle = 3 \times 10^{-26}\ \rm cm^{3}\ s^{-1} $, $ B = 3\ \rm \mu G $, and $ \nu = 30\ \rm GHz $. The spikes are characterized by $ \gamma_{\mathrm{spike}} = 7/3 $, $ r_{\mathrm{sat}} = r_{\mathrm{Sch}} $ (left panel), $ r_{\mathrm{sat}} = r_{\mathrm{sat}}^{\mathrm{ann}} $ (right panel), and different radii. The blue vertically hatched area represents the additional uncertainty due to diffusion and the unknown cross section, bracketed by the flux for $ \left\langle \sigma v \right\rangle = 3 \times 10^{-27}\ \rm cm^{3}\ s^{-1} $ and $ \left\langle \sigma v \right\rangle = 3 \times 10^{-25}\ \rm cm^{3}\ s^{-1} $.}
\end{figure*}

Complementary information on the DM profile can be gained by looking at the very inner region of the Galaxy. Hence, we shall now study the synchrotron emission at angular scales down to a few $ \rm \mu as $, in the framework of a futuristic telescope with $ \rm \mu as $ resolution at both radio and millimetre frequencies. In the near future, such a high resolution may only be attained by the Event Horizon Telescope network \cite{EHT}, for higher frequencies, typically of the order of $ 400\ \rm GHz $. 

By looking at these very small scales, one expects to be more sensitive to the characteristics of the spike. Our estimates of the fluxes below $ 0.1^{\circ} $ are given in Fig.~\ref{flux spike zoom} (left panel), for spiky profiles of $R_{\rm{spike}}= 0.1,1,10$ pc and a saturation radius $ r_{\mathrm{sat}} = r_{\mathrm{Sch}} $. For comparison we also display the flux for the NFW DM halo profile. As one expects, the fluxes associated with spiky profiles become extremely large toward the GC. The reason is that for such values of $ r_{\mathrm{sat}} $, the spike becomes so steep toward the center that diffusion becomes negligible below $\simeq 100 \ \rm \mu as $. Hence, a large portion of the electrons stay confined in the inner part and do not diffuse outside the center. Above $\simeq 100\ \rm \mu as $ diffusion is important, so the synchrotron emission is smeared out accordingly.
 
Assuming $ r_{\mathrm{sat}} = r_{\mathrm{sat}}^{\mathrm{ann}} $ (cf.~Fig.~\ref{flux spike zoom}, right panel) leads to very different fluxes: not only do they reach a plateau below $b \sim 1\ \rm \mu as$, but also the corresponding value is much smaller than in the $ r_{\mathrm{sat}} = r_{\mathrm{Sch}} $ case. The main explanation is that the DM distribution has a much larger core in this case, so the number of electrons and positrons injected by the DM is constant at distance $r< r_{\mathrm{sat}}$ and is also much smaller than when one assumes $ r_{\mathrm{sat}} = r_{\mathrm{Sch}} $. Diffusion is more effective then, and as a result the synchrotron flux is much smaller when $ r_{\mathrm{sat}} = r_{\mathrm{sat}}^{\mathrm{ann}} $ than in the $ r_{\mathrm{sat}} = r_{\mathrm{Sch}} $ case. These results therefore could be used to constrain the saturation radius.

\subsubsection{Combining small and large scales}

On the one hand, one can determine the size of the saturation radius by using the value of the flux below $ 10^{-3} \, \mathrm{as} $. On the other hand, one can infer the size of the spike by studying the morphology at $ 0.1^{\circ} \lesssim b \lesssim 10^{\circ} $ scales. As the size of the spike enters the expression of the saturation radius, the combination of observations from small to large scales should provide us with a consistent picture of the DM inner profile, potentially also pointing toward the value of the cross section if dark matter is indeed made of annihilating particles. These measurements could therefore be used to verify or infer the nature of dark matter.
 
Note that to draw our conclusions we used the canonical value of $ \left\langle \sigma v \right\rangle = 3 \times 10^{-26}\ \rm cm^{3}\ s^{-1} $. To test the robustness of our claim, we now estimate the uncertainty on the flux due to the lack of determination of the cross section. We therefore consider two values $ \left\langle \sigma v \right\rangle = 3 \times 10^{-27}\ \rm cm^{3}\ s^{-1} $ and $ \left\langle \sigma v \right\rangle = 3 \times 10^{-25}\ \rm cm^{3}\ s^{-1} $ and assume the existence of a regeneration mechanism for DM particles when $ \left\langle \sigma v \right\rangle > 3 \times 10^{-26}\ \rm cm^{3}\ s^{-1} $ \cite{regeneration}. 

The uncertainty on the flux due to both uncertainties in diffusion and the broader range for the annihilation cross section is represented by the blue vertically hatched area in Fig.~\ref{flux spike zoom}. From this figure we can see that the morphology inferred by using $ \left\langle \sigma v \right\rangle = 3 \times 10^{-26}\ \rm cm^{3}\ s^{-1} $ is unchanged when the cross section is increased or decreased. Thus, changing the cross section only affects the normalization of the flux.

In principle not knowing the cross section could lead to a misinterpretation of the spike characteristics: assuming the canonical cross section, one could deduce the wrong values for $ R_{\mathrm{spike}} $ or $ r_{\mathrm{sat}}$. However, since one can determine $ R_{\mathrm{spike}} $ using the data at high latitudes and the morphology of the emission, the only possible source of degeneracy is between $\left\langle \sigma v \right\rangle$ and $ r_{\mathrm{sat}}$. In the case of annihilating DM, this should not be a problem as both quantities are related. This is more problematic if there is no evidence that DM is annihilating, but one would not expect any anomalous synchrotron emission from the GC (unless DM is decaying, in which case the decay rate and $ r_{\mathrm{sat}}$ should also be related). 

As for distinguishing decaying from annihilating  DM, for a given density profile, the morphology of the emission is different in both cases, as shown in Refs.~\cite{ascasibar,BoehmDelahayeSilk}. One can therefore in principle discriminate between annihilating and decaying DM, but repeating a similar analysis for decaying DM is beyond the scope of our paper. Annihilating and decaying DM are degenerate in terms of morphology only if the DM profile is twice as steep for decaying DM as for annihilating DM. However, in this work, we focus on the profile of annihilating DM, typically a spike with $ \gamma_{\mathrm{spike}} = 7/3 $. Mimicking the morphology of the resulting emission with decaying DM would require a DM profile with a power-law index of the order of 5, which is unrealistic.

\subsubsection{Other values of the spike index}

So far, we only have considered spikes with a power-law index $ \gamma_{\mathrm{spike}} = 7/3 $ that corresponds to the prescription given in Ref.~\cite{Gondolo}. However, WIMPs scattering off stars in a dense star cluster at the GC may lead to shallower DM spikes with $ \gamma_{\mathrm{spike}} = 3/2 $ \cite{Gnedin}. 

Assuming $ \gamma_{\mathrm{spike}} = 3/2 $ and $ r_{\mathrm{sat}} = r_{\mathrm{sat}}^{\mathrm{ann}} $, we obtain, however, the same flux as for a NFW profile. Our result is independent of the size of the spike and the latitude because the number of electrons injected in the center is small enough for diffusion to be efficient. Said differently, diffusion washes out the signature of a spike when the index is $\gamma_{\mathrm{spike}} = 3/2$ and the saturation radius is fixed by the annihilation cross section. 

For $ r_{\mathrm{sat}} = r_{\mathrm{Sch}}$, diffusion is only efficient above $ 100\ \rm \mu as $. Below $ 100\ \rm \mu as$, the flux still shows evidence for a spike. Therefore, such (not too steep) profiles could be seen by making observations below $ 100\ \rm \mu as $ if the saturation radius were extremely small indeed.

\subsection{Impact of the magnetic field}

\begin{figure*}[tbp]
\centering
\includegraphics[scale=0.41]{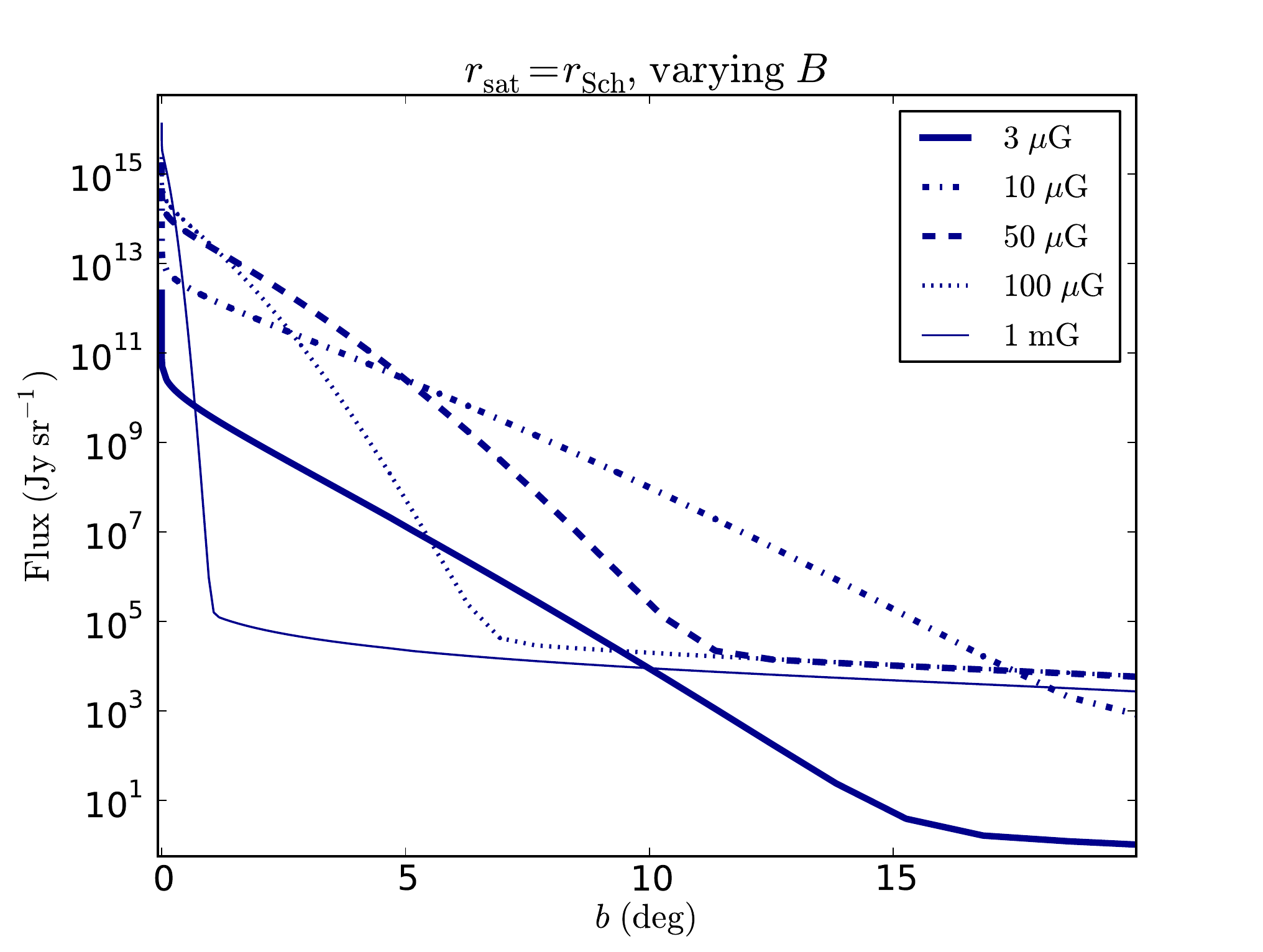}\includegraphics[scale=0.41]{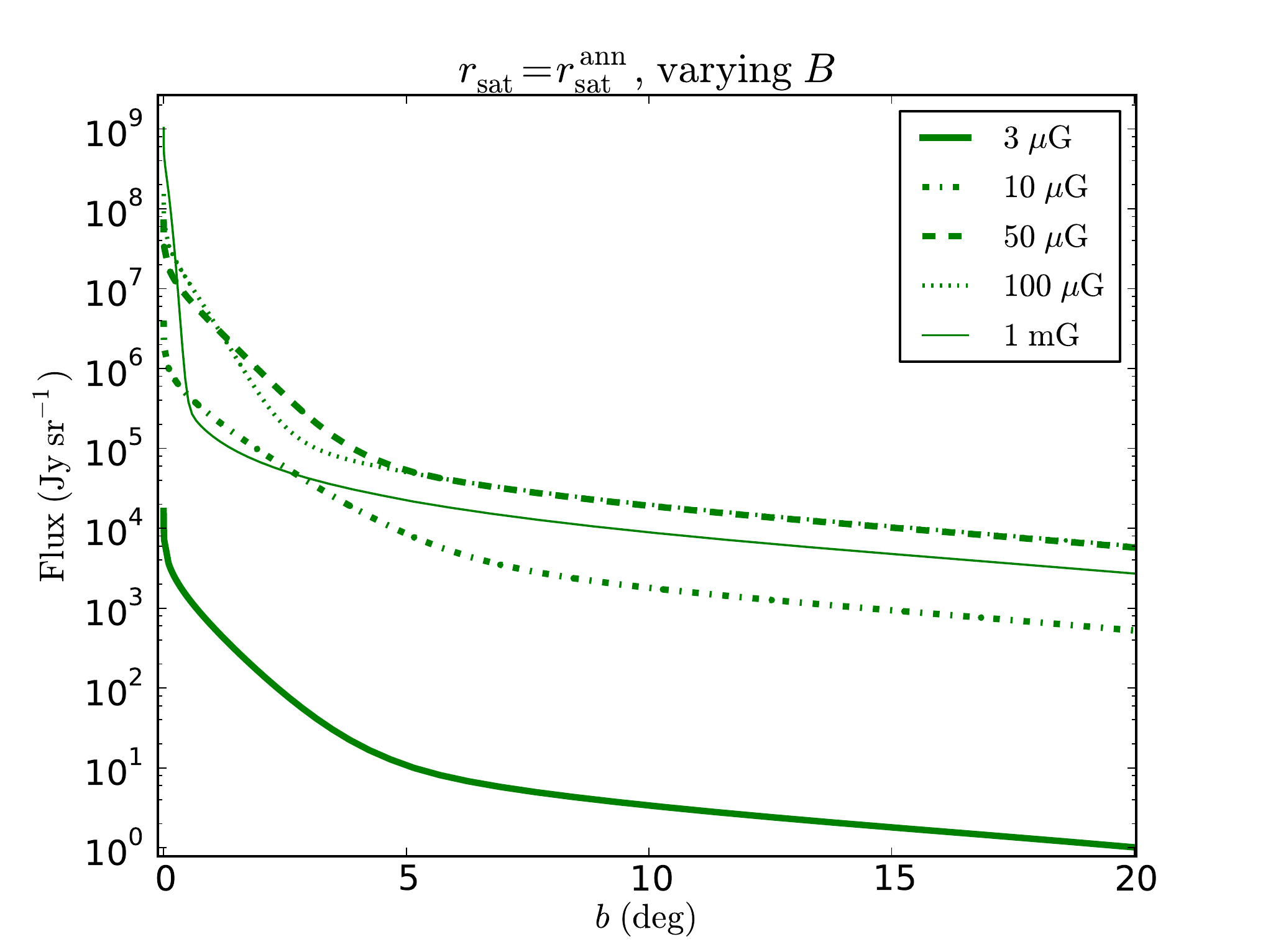} 
\caption{\label{flux_30GHz_10GeV_3e-26_latitude_B}Synchrotron flux as a function of latitude $ b $, for $ 10 \ \rm GeV $ DM particles, $ \left\langle \sigma v \right\rangle = 3 \times 10^{-26}\ \rm cm^{3}\ s^{-1} $, $ \nu = 30\ \rm GHz $, for a spike with $ \gamma_{\mathrm{spike}} = 7/3 $, $ r_{\mathrm{sat}} = r_{\mathrm{Sch}} $ (left panel), $ r_{\mathrm{sat}} = r_{\mathrm{sat}}^{\mathrm{ann}} $ (right panel), and for 5 values of the magnetic field between $ 3\ \rm \mu G $ and $ 1\ \rm mG $. The MED set of propagation parameters is used.}
\end{figure*}

\begin{figure}[tbp]
\centering
\includegraphics[scale=0.41]{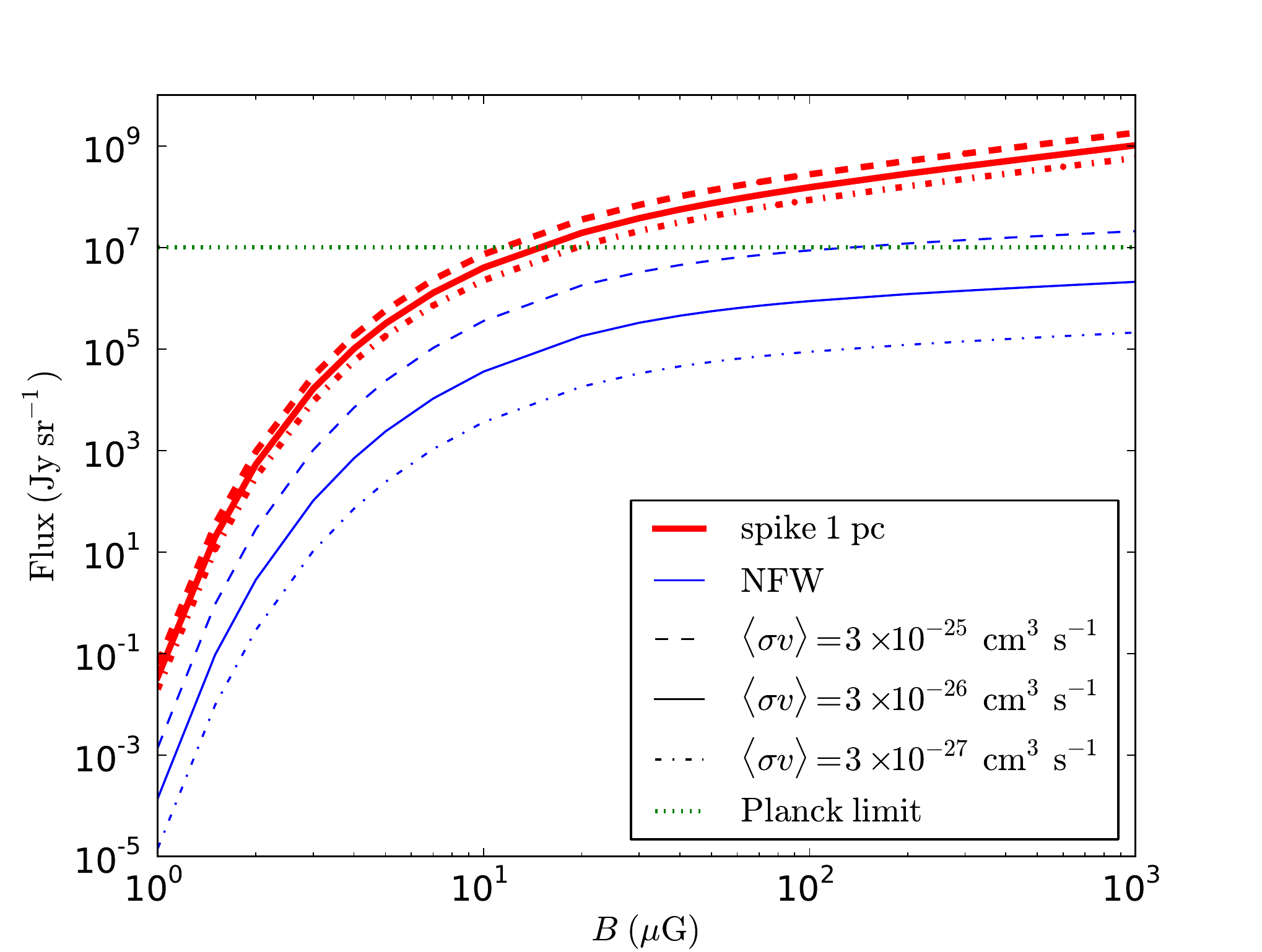} 
\caption{\label{flux_vs_B}Synchrotron flux from the direction of the GC ($ l = b = 0^{\circ} $) as a function of the magnetic field intensity, for $ 10 \ \rm GeV $ DM particles, $ \nu = 30\ \rm GHz $, and for the NFW profile (blue thin lines) and the NFW+spike profile with $ \gamma_{\mathrm{spike}} = 7/3 $, $ R_{\mathrm{spike}} = 1\ \rm pc $, and $ r_{\mathrm{sat}} = r_{\mathrm{sat}}^{\mathrm{ann}} $ (red thick lines). The MED set of propagation parameters is used. The green dotted line represents the limit on the flux given by Planck.}
\end{figure}

\begin{figure}[tbp]
\centering 
\includegraphics[scale=0.41]{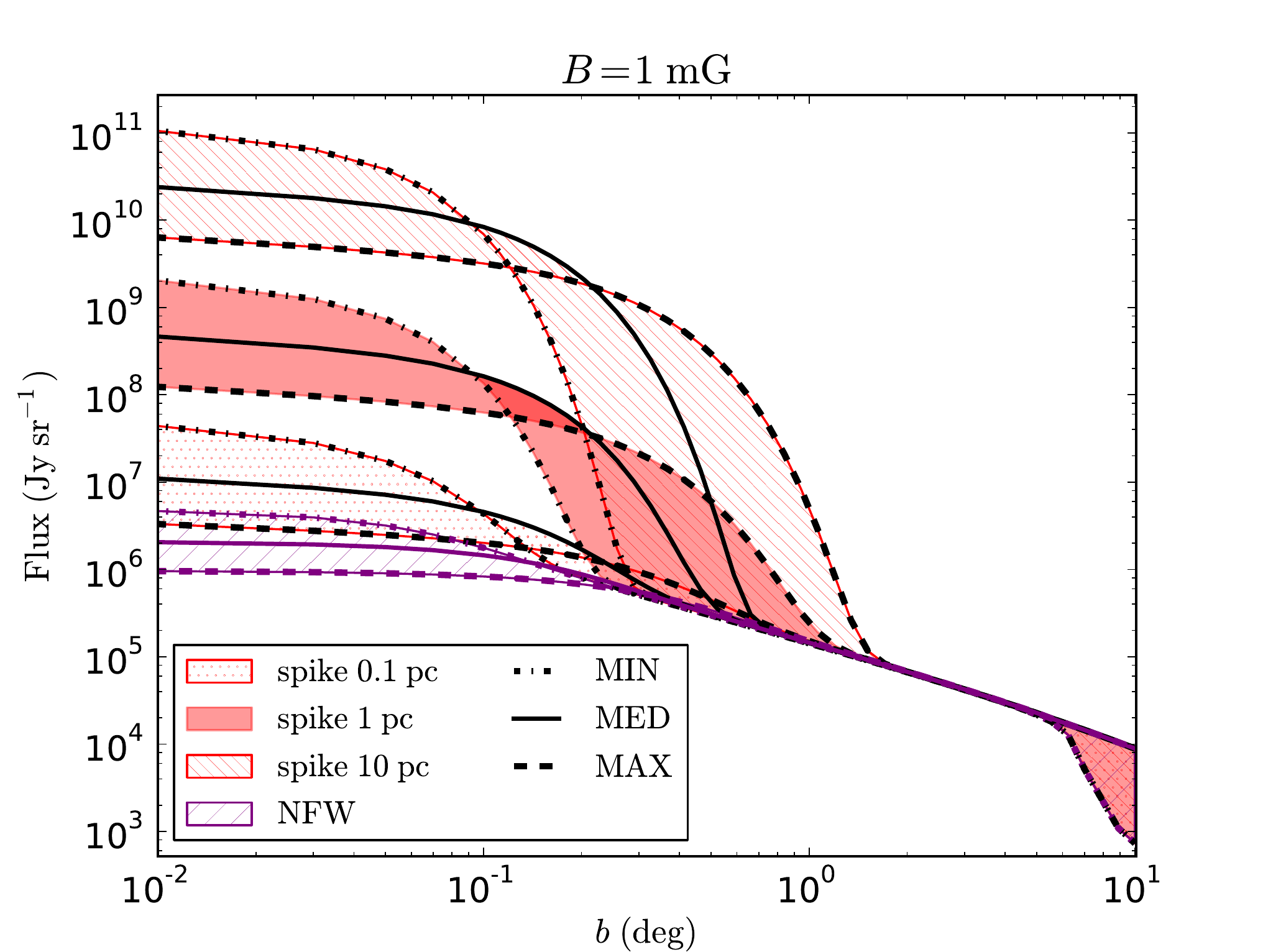} 
\caption{\label{flux 3 spikes 1mG}Synchrotron flux as a function of latitude $ b $, for $ 10 \ \rm GeV $ DM particles, $ \left\langle \sigma v \right\rangle = 3 \times 10^{-26}\ \rm cm^{3}\ s^{-1} $, $ B = 1\ \rm mG $, and $ \nu = 30\ \rm GHz $, for spikes with $ \gamma_{\mathrm{spike}} = 7/3 $, $ r_{\mathrm{sat}} = r_{\mathrm{sat}}^{\mathrm{ann}} $, and different radii.}
\end{figure}

We can now study the influence of the magnetic field intensity on the flux. To avoid possible degeneracies between the impact of a spike and spatial variations of the magnetic field, we will consider a constant field intensity over the whole diffusion zone. There is no established value of the magnetic field around Sgr A*. Throughout our study we have used $ B = 3\ \rm \mu G$, which is the expected value at large angular scales, but a recent study suggests that $B$ could actually be greater than $ 1\ \rm mG $ \cite{Magnetic_field_nature} in the GC. To test the robustness of our conclusions, we now investigate the impact of the magnetic field intensity on the morphology of the synchrotron emission. Our results are shown in Fig.~\ref{flux_30GHz_10GeV_3e-26_latitude_B}, where we see that increasing the magnetic field from $ 3\ \rm \mu G$ to $1\ \rm mG $ can significantly affect both the normalization and the morphology of the signal on scales of a few degrees. This is true in fact whether we consider $ r_{\mathrm{sat}} = r_{\mathrm{sat}}^{\mathrm{ann}}$ or $ r_{\mathrm{sat}} = r_{\mathrm{Sch}} $.

This can be understood as follows: the synchrotron flux is the integral over the energy of the halo function times the ratio of the synchrotron power to the losses. All these quantities depend on the magnetic field but in different ways:

(i) The synchrotron power is proportional to the magnetic field as $P_{\mathrm {syn}}(E) \propto B F_\mathrm{i}(x)$.

(ii) The losses, being in the first approximation the sum of IC and synchrotron contributions, are dominated by one or the other depending on the value of the magnetic field; they are either almost independent of $ B $ when IC losses dominate or proportional to the magnetic field squared when the synchrotron losses are dominant.

(iii) The halo function $\tilde{I}$ is not directly proportional to the magnetic field. However, the larger the magnetic field, the more confined the electrons, so when the magnetic field increases, the calculation of $\tilde{I}$ becomes essentially dominated by the very small values of the propagation length ($\lambda_{\mathrm D} \rightarrow 0$). The halo function is therefore related to the magnetic field in a nontrivial way. 

In the regime where the IC processes are the dominant contribution to the loss term, the dependence of the flux on the magnetic field mostly arises through the expression of the synchrotron power. At $30 \ \rm GHz$ and for $B \in [3,10] \, \mathrm{\mu G}$, we find that $F_\mathrm{i}(x) \propto B^{p}$ so $P_{\rm syn}(E) \propto B^{p+1}$ with $p\approx 4$, depending on the exact value of the energy. An increase in $B$ thus induces a global increase in the flux, as can be seen in Fig.~\ref{flux_30GHz_10GeV_3e-26_latitude_B} by comparing our predictions for $B=3 \, \mathrm{\mu G}$ and $B=10 \, \mathrm{\mu G}$.

In the intermediate regime where $ B \in [10,100]\ \rm \mu G$, IC and synchrotron losses are about the same order of magnitude, so the dependence of the flux on $B$ is more complex. It grows from $B^0$ to $B^2$. As a result, at high latitude where $ \tilde{I}$ is independent of $E$ and $B$, the dependence of $\int dE P_{\mathrm{syn}}(E) \tilde{I}/b(E)$ with the magnetic field decreases from $B^5$ to $1/\sqrt{B}$, while at low latitude the electrons are more and more confined as $B$ increases, so the morphology of the emission strongly depends on $B$. Finally in the regime where $ B \gtrsim 100\ \rm \mu G$, the synchrotron losses are dominant so $\Phi_{\nu} \propto 1/\sqrt{B}$ at high latitude.

For a given value of the magnetic field, the flux as a function of the latitude follows the behavior of the halo function, which describes the outcome of the diffusion in terms of confinement. The latitude at which the flux reaches its lower value is determined by the magnetic field. The stronger $B$, the smaller the confinement region and the earlier the flux reaches its lower plateau in terms of latitude. The plateau feature is more pronounced when $ r_{\mathrm{sat}} = r_{\mathrm{Sch}} $ than when $ r_{\mathrm{sat}} = r_{\mathrm{sat}}^{\mathrm{ann}}$ because the number density of electrons is larger in the GC for $ r_{\mathrm{sat}} = r_{\mathrm{Sch}} $, so the effect of confinement is more pronounced (as can be seen by comparing the left and right panels of Fig.~\ref{flux_30GHz_10GeV_3e-26_latitude_B}). 

We can now focus on the critical influence of the magnetic field on the normalization of the flux. As shown in Fig.~\ref{flux_vs_B} (and by comparing Fig.~\ref{flux 3 spikes}, left panel, to Fig.~\ref{flux 3 spikes 1mG}), the flux varies by more than 4 orders of magnitude between $ 3\ \rm \mu G $ and $ 1\ \rm mG $. Consequently, the magnetic field has a huge impact on the constraints that one can set on the existence of a spike and its size. Large values of the magnetic field lead to a large flux and thus potentially offer a scope for detectability of a steep inner profile.

\subsection{Observability by Planck}

We can now tackle the chances to probe the existence of a spike by the Planck experiment. Using the results from the Planck collaboration \cite{Planck_results}, we estimate the total flux at $ 30\ \rm GHz $ from the GC to be of the order of $ 10^{7}\ \rm Jy\ sr^{-1} $. Since we do not take into account the resolution of the detector, comparing our estimates of the flux with this value only provides us with an indication of the synchrotron limit on these scenarios rather than a strict constraint. However, such a value turns out to be very useful in order to determine the ability of the Planck experiment to probe the existence of a spike. 

From Fig.~\ref{flux spike zoom}, left panel, we see that any spike with an extremely small saturation radius $ r_{\mathrm{sat}} = r_{\mathrm{Sch}} $ actually predicts a much larger flux than what has been observed by the Planck collaboration. Therefore, such profiles are likely to be excluded (especially since we used $B = 3\ \rm \mu G$, which is a conservative value). Inspecting the right panel of Fig.~\ref{flux spike zoom} shows that spikes with a saturation radius of $ r_{\mathrm{sat}} = r_{\mathrm{sat}}^{\mathrm{ann}} $ predict fluxes below the Planck limit, thus indicating that Planck may not be able to set meaningful constraints. However, these results were obtained by assuming $B= 3\ \rm \mu G$ and the canonical value of the annihilation cross section. Taking $ B \gtrsim {\cal{O}}(10)\ \rm \mu G $ (or a larger cross section value if one also assumes a regeneration mechanism \cite{regeneration}) increases these fluxes by several orders of magnitude and typically implies that they exceed the Planck limit; cf.~Fig.~\ref{flux_vs_B}. Hence, if one assumes a reasonable value of the magnetic field in the GC, we find that Planck is likely to be able to probe these spikes. 

This is illustrated in Fig.~\ref{flux 3 spikes 1mG}, where we display the synchrotron flux for a very large $B$ value ($ B = 1\ \rm mG $) and the same parameters as in Fig.~\ref{flux 3 spikes} (left panel). As one can readily see, spikes with radii 1 and 10 pc are excluded as their fluxes exceed the Planck limit. Therefore, we conclude that the Planck experiment has the ability to constrain the presence of spiky DM halo profiles and discriminate between spikes of different sizes if there is a strong magnetic field in the GC. 

The same types of conclusions hold for heavy (800 GeV) DM particles. At 857 GHz, the Planck limit on the emission from the GC is, however, of the order of $ 10^{9}\ \rm Jy\ sr^{-1} $ \cite{Planck_results}. Assuming $ B = 1\ \rm mG $ and $ r_{\mathrm{sat}} = r_{\mathrm{sat}}^{\mathrm{ann}} $, we expect the synchrotron flux (for $ \left\langle \sigma v \right\rangle = 3 \times 10^{-26}\ \rm cm^3\ s^{-1} $) to be about $ 10^{5}\ \rm Jy\ sr^{-1} $ (for MED). This is actually below the Planck limit, and so the presence of a spike would be difficult to assess in this case. However, a smaller saturation radius or an even larger magnetic field would increase the flux.

Note that there could be additional constraints other than Planck on 10 GeV DM. For large values of the magnetic field, 10 GeV DM particles overproduce the synchrotron emission with respect to Sgr A* at radio frequencies ($ 300-400\ \rm MHz $) and are therefore likely to be excluded \cite{BEnssS,BEnssS2}. One important caveat, however, is that at such low frequencies one must take into account the effects of advection and self-absorption of the synchrotron emission \cite{multi-wavelength}, which were neglected in Ref.~\cite{BEnssS,BEnssS2}. These effects could reduce the radio flux and potentially weaken the radio constraints. Since such advection and self-absorption effects can be safely neglected at $ 30\ \rm GHz$, using Planck data to constrain 10 GeV DM and the inner profile should provide us with a more robust method, although the foreground emission could then be problematic. 

In our analysis we have chosen a constant magnetic field over the whole Galaxy. Better modelling of this field across the Galaxy would improve the analysis, but this is beyond the scope of this paper. Also we remark that our assumption of a very large (and constant) magnetic field is not realistic as one expects $B\sim 3 \, \mathrm{\mu G}$ far away from the center. However, due to the confinement effect associated with large values of $ B $, our conclusions should remain unchanged in that specific case.

\section{Conclusion and perspectives}

In this work, we have investigated whether it is possible to probe the DM energy distribution in the inner part of the Galaxy using synchrotron emission. We have focused on light (10 GeV) DM annihilating into $e^+ e^-$ but we also have investigated the case of heavy (800 GeV) DM. We have considered several DM halo profiles with different behaviors toward the GC, namely, NFW, NFW$+$spike with index $\gamma_{\rm{spike}} \sim 7/3$ and several sizes for the spike ($R_{\rm{spike}} = 0.1,1,10$ pc). We also have assumed that the energy density eventually reaches a plateau at scales smaller than a saturation scale $r_{\rm{sat}}$, which we have chosen to be either determined by the annihilation cross section ($r_{\mathrm{sat}} = r_{\mathrm{sat}}^{\mathrm{ann}} $) or independent of the annihilation cross section and given by the Schwarzschild radius ($ r_{\mathrm{sat}} = r_{\mathrm{Sch}} $).

The standard propagation techniques that exist in the literature do not enable one to account for the increase in the electron number density close to the GC. We have therefore modified the standard treatment of cosmic ray propagation to account for a steep energy injection profile in the GC. Armed with the calculation of the electron and positron energy distribution after propagation, we have been able to study the morphology of the synchrotron emission that is expected from annihilating DM candidates. 

Our main conclusions are the following: first, we have shown that the size of the spike $R_{\rm{spike}}$ leaves an imprint on the synchrotron flux at degree scales, and, second, the saturation radius $r_{\mathrm{sat}} $ can be inferred by zooming in on the GC. This second point prefers an instrument with very good resolution ($\mu \rm as$), although this is not crucial. We thus find that the combination of small and large scales could enable one to probe the existence of a spiky DM halo distribution in the inner Galaxy. We also observe that using Planck data only could enable one to probe spikes of size greater than 1 pc, provided that the magnetic field is larger than $\sim 20\ \rm \mu G$ in the inner center and that the cross section is not too small. One can of course extend this analysis to other annihilation channels, but this is beyond the scope of this paper.

We note also that the Event Horizon Telescope will be able to probe a DM spike (and determine $r_{\mathrm{sat}} $) around the much more massive black hole in M87. This is particularly important because the spike profile may be strongly affected by dynamical interactions with stars as argued in Ref.~\cite{Gnedin}. While this effect, however, is probably important for our GC, the effects of relaxation are unimportant for the case of M87, where the dynamical relaxation time in the core is very much longer: $ 10^5 $ Gyr vs several Gyr for our GC. Hence, the initial steep DM spike should be preserved. We will discuss the potential of observations of the center of M87 in a future paper.

Finally, in addition to probing the existence of a spike in the inner Galaxy, another application of this work could be to improve the foreground modelling, in particular for Planck. Adding the emission induced by a DM spike to the astrophysical component might allow one to jointly constrain the properties of the spike and refine the foreground models.

\begin{acknowledgements}
We would like to thank Timur Delahaye for fruitful discussions. This research has been supported at IAP by the ERC Project No.~267117 (DARK) hosted by Universit\'e Pierre et Marie Curie (UPMC) - Paris 6 and at JHU by NSF Grant No.~OIA-1124403. This work has been also supported in part by Ecole Normale Sup\'{e}rieure de Lyon, UPMC and STFC. 
\end{acknowledgements}

\section*{Appendix: Coordinate systems}

\newpage

\bibliographystyle{h-physrev} 
\bibliography{biblio_thesis}

\end{document}